\documentclass[twocolumn,tighten,times]{aastex63}

\hypersetup{colorlinks=true,linkcolor=blue,citecolor=blue,urlcolor=blue,}

\shorttitle{\sc {\MgII} Voigt Profile Models}
\shortauthors{\sc Churchill \etal}

\input{iondefs.sty}

\begin{document}

\title{M{\MakeLowercase g}\kern 0.1em{\small II} Absorbers in High-Resolution Quasar Spectra. I. Voigt Profile Models}

\author[0000-0002-9125-8159]{Christopher W. Churchill}
\affiliation{Department of Astronomy, New Mexico State University, Las Cruces, NM 88003, USA}

\author[0000-0002-6846-6582]{Jessica L. Evans}
\affiliation{Department of Astronomy, New Mexico State University, Las Cruces, NM 88003, USA}

\author[0000-0002-6434-4684]{Bryson Stemock}
\affiliation{Department of Astronomy, New Mexico State University, Las Cruces, NM 88003, USA}

\author[0000-0003-2377-8352]{Nikole M. Nielsen}
\affiliation{Centre for Astrophysics and Supercomputing, Swinburne University of Technology, Hawthorn, Victoria 3122, Australia}
\affiliation{ARC Center for Excellence for All Sky Astrophysics in 3-Dimensions (ASTRO-3D)}

\author[0000-0003-1362-9302]{Glenn G. Kacprzak}
\affiliation{Centre for Astrophysics and Supercomputing, Swinburne University of Technology, Hawthorn, Victoria 3122, Australia}
\affiliation{ARC Center for Excellence for All Sky Astrophysics in 3-Dimensions (ASTRO-3D)}

\author[0000-0002-7040-5489]{Michael T. Murphy}
\affiliation{Centre for Astrophysics and Supercomputing, Swinburne University of Technology, Hawthorn, Victoria 3122, Australia}


\begin{abstract}
We present the Voigt profile (VP) models, column densities, Doppler $b$ parameters,  kinematics, and distribution of components for 422 {\MgII} absorbers found in a survey of 249 HIRES and UVES quasar spectra. The equivalent width range of the sample is $0.006 \leq W_r(2796) \leq 6.23$ {\AA} and the redshift range is $0.19 \leq z \leq 2.55$, with a mean of $\langle z  \rangle = 1.18$. Based on historical precedent, we classified 180 absorbers as weak systems ($W_r(2796) < 0.3$ {\AA}) and 242 as strong systems ($W_r(2796) \geq 0.3$  {\AA}).  Assuming a minimum number of significant components per system, the VP fitting, yielded a total of 2,989 components, with an average of 2.7 and 10.3 components found for the weak and strong {\MgII} subsamples, respectively.  The VP component line density for the full sample is $8.62 \pm 0.23$  clouds {\AA}$^{-1}$. The distribution  of VP component column density over the range $12.4 \leq \log N({\MgII}) \leq 17.0$  cm$^{-2}$ is well modeled with a power-law slope of $-1.45\pm0.01$.  The median Doppler $b$ parameters are $4.5\pm3.5$~{\kms}, $6.0\pm4.5$~{\kms}, and $5.7\pm4.4$~{\kms} for the weak, strong, and full samples. We modeled the probability of component velocity splitting (the  two-point velocity correlation function, TPCF) of our full sample using a three-component  composite Gaussian function. Our resulting velocity dispersions are $\sigma_1 =  25.4$~{\kms}, $\sigma_2 = 68.7$~{\kms}, and $\sigma_3 = 207.1$~{\kms}. These data provide  an excellent database for studying the cosmic evolution of {\MgII} absorber kinematic  evolution.\\
\end{abstract}


\section{Introduction}
\label{sec:intro}

Following the first suggestions that the ``forest" of {\Lya} absorption lines in the spectra of quasars implied a ubiquitous yet porous intergalactic gaseous medium \citep[e.g.,][]{bergeron70, arons72} and the hypothesis that the rich array of narrow metal-absorption lines arise from extended gaseous halos around galaxies \citep[e.g.,][]{bahcall69}, the study of quasar absorption lines has developed into a powerful experimental tool for characterizing the properties of the intergalactic medium (IGM) and the circumgalactic medium (CGM). It is through the quantified analysis of absorption lines that we theorize {\it how\/} galaxies interact with a gaseous cosmic web and partake in a ``baryon cycle" in which gas cycles into, out of, and through galaxies. This baryon cycle is arguably one of the most important physical processes governing the evolution of the observable universe of stars, galaxies, and cosmic chemical enrichment. 

From the time the first quasar spectra were of high enough resolution to yield line profile shapes, clear velocity splitting was observed \citep[e.g.,][]{bahcall75, boksy75}; it was quickly understood that the kinematic, chemical, and ionization conditions of high-redshift gaseous systems could be studied in detail. Voigt profile (VP) fitting of the absorption profiles was immediately employed \citep[e.g.,][]{morton72a, morton72b, boksy79}, as it conveniently allows a well-posed model of the data that naturally accounts for the instrumental line spread function and multiple absorption components, and yields the individual component column densities of the absorbing ions and their thermal broadening Doppler $b$ parameters. The central limitation to the VP model is the assumption that each component is a spatially isolated isothermal ``cloud". Nonetheless, within the confines of these assumptions, ionization modeling based on VP fitting parameters was soon applied  \citep[e.g.][]{bergeron86, steidel90, verner90}, and new insights into the IGM and CGM were garnered.

Over the last several decades, VP fitting of quasar absorption lines has been a key part of transforming and advancing our cosmic perspective of gaseous structures throughout the universe. Constraints on the redshift clustering and column density and temperature distributions of the IGM have been obtained using VP fitting of {\Lya} forest lines \citep[e.g.,][]{morton72a, hu95, lu96, kirkman97, kim07, misawa07, danforth10, rudie12, kim13, hiss18, garzilli20}. As absorption systems having $\log N({\HI}) \geq 17.2$ {\cmsq}, i.e., the so-called Lyman Limit, sub-damped, and damped {\Lya} systems, are associated with a wide array of metal lines exhibiting kinematic complexity, VP fitting has been instrumental in characterizing the physical conditions of the many astrophysical environments they probe \citep[e.g.,][]{peroux06, rao06, meiring08, prochter10, lehner14, prochaska15, lehner16, lehner18}. VP fitting to these systems has even been instrumental in constraining key cosmological parameters, such as the cosmic baryon density \citep[e.g.,][]{burles98, tytler99} and possible cosmic evolution of fundamental constants, such as the fine structure constant \citep[e.g.,][]{webb99, murphy17}.

Constraints on the spectral energy distribution of the ionizing background radiation and cosmic mass density, as well as the kinematic, chemical, and ionization conditions and the evolution in these properties in both the CGM and IGM, have been studied using VP fitting to {\CIV}-selected absorbers \citep[e.g.,][]{morton72b, rauch96, petitjean94, songaila98, kim02, simcoe04, ryan-webbet06, becker09, boksy15, cooper19, manuwal19}. The high-ionization CGM and the chemical enrichment and physical conditions of the high-ionization IGM have also been extensively studied using the VP methodology applied to {\OVI}-selected and {\OVI}+{\Lya} absorbers \cite[e.g.,][]{simcoe02, simcoe04, danforth06, tripp08, muzahid12, johnson13, werk13, mathes14, savage14, muzahid15, pointon19}, including those exhibiting {\NeVIII} absorption \citep[e.g.,][]{savage05}. Similarly, the kinematic, chemical, and ionization conditions of {\MgII}-selected absorbers, which are primarily associated with the low-ionization CGM, have also been studied in detail using VP fitting \citep[e.g.,][also see \citealp{cooper19}]{churchill97, rigby02, churchill03, lynch07, narayanan08, evans-phd}.  

In the modern era, it is well known that the parameters derived from VP modeling for a given sample of absorption line systems cannot be unique, as the fitting is not based on strictly objective criteria. Human subjectivity plays a role such that different humans will adopt different VP models for the same absorption systems even when they employ the same VP fitting software. And yet, many VP fitting routines have been developed and applied to quasar absorption line systems \citep[e.g.,][]{vidal77, welty91, carswell91, fontana95, mar95, churchill97, foreman13, howarth15, bainbridge17a, galkwad17, liang17, krogager18, cooke19}. Scores of studies of {\Lya} forest, Lyman limit, damped {\Lya}, {\OVI}, {\CIV}, and {\MgII}-selected systems have each employed one VP fitting code or another, and this also affects the reproducibility of VP models \citep[however, the most commonly used fitting routine is {\sc VPfit},][]{carswell14}

Though efforts are being pioneered to mitigate human subjectivity \citep[e.g.,][]{bainbridge17a, bainbridge17b}, there may always be differences in fitting approach due to differing science objectives. Examples might include whether to model the absorption profiles with the minimum number of components \citep[e.g.,][]{churchill97, churchill03}, or with the number required to minimize all pixel residuals below some minimum \citep[e.g.,][]{murphy01, murphy17, bainbridge17a}, or how to segregate components between the transitions of high- and low-ionization species \citep[e.g.,][]{simcoe06, muzahid15, rudie19}.  

Although the resonant {\MgIIdblt} fine-structure absorption lines are among the most common transitions populating quasar spectra \citep[e.g.,][]{lanzetta87, steidel92, nestor05, prochter06, zhu13}, the statistics and distribution of their kinematics from VP fitting has been documented only in relatively small numbers (less than 50) and for relatively low redshifts, $z \leq 1.5$ \citep{churchill97, churchill03}. As such, the kinematics of low-ionization, chemically enriched CGM gas has not been characterized using VP derived parameters probing the epoch known as ``Cosmic Noon" ($z\! \sim\! 2$--6), when  the global star formation rate density of the universe rose toward its peak  \citep[e.g.,][]{madau14}, stellar-driven outflows from galaxies became ubiquitous \citep[e.g.,][]{rupke18} and the predicted rate of gas accretion into galaxies reached its cosmic peak \citep[e.g.,][]{freeke11b}.  Simulations further suggest that it is the drop in this accretion rate that is responsible for the decline in the star formation density following Cosmic Noon \citep[e.g.,][]{freeke11a}.  


The cosmic evolution of outflows and accretion through the CGM is expected to be reflected in the absorption properties of {\MgII} absorbers.  Indeed, both the observed evolution in the comic incidence (redshift path density) of {\MgII} absorbers \citep[e.g.,][]{zhu13} and the relative frequency of higher equivalent width systems \citep[e.g.,][]{matejek12} traces the evolution of the global star formation density. Studies of {\MgII} absorption in relation to their host galaxies at have yielded a preponderance of observational evidence for outflows associated with star formation \citep[e.g.,][]{bouche06, tremonti07, martin09, weiner09,  noterdaeme10, kacprzak10, kacprzak14, rubin10, coil11, nestor11, martin12, noterdaeme12, krogager13, peroux13, crighton15, nielsen15, nielsen16, lan18, schroetter19, zabl20} 
as well as evidence for accretion \citep[e.g.,][]{steidel02, kacprzak10, peroux13, ho17, kacprzak17, martin19, zabl19}.

With the modern capabilities of the VLT/MUSE instrument \citep{bacon04}, surveys such as MAGG \citep[][]{lofthouse20} at $z\simeq 3.5$ will soon be yielding numerous absorber-galaxy pairs at times preceding Cosmic Noon \cite[also see][]{mackenzie19, gonzalo20}.  Similarly, with the bluer sensitivity of the KCWI instrument \citep{morrissey18}, we are characterizing the properties of galaxies at Cosmic Noon for which the CGM kinematics is studied via high-resolution {\MgII} absorption \citep[e.g.,][]{nielsen19, nielsen20}.  Motivated by (1) indications that CGM kinematic evolution is occurring between Cosmic Noon and the present epoch, (2) that this evolution likely traces the global star formation density and can provide insights into the physics of the baryon cycle, and (3) the growing samples of {\MgII} absorption-galaxy pairs at Cosmic Noon, we have undertaken VP profile fitting of a sample of several hundred {\MgII} absorption systems spanning equivalent widths $0.006 \leq W_r(2796) \leq 4.23$~{\AA} over the redshift range $0.2 \leq z \leq 2.6$.

In this paper, we present and discuss the results of our VP modeling.  In Section~\ref{sec:data} we present the sample of quasar spectra and describe our methods of preparing the data into science-ready form, identifying {\MgII}-selected absorbers, and quantifying absorption properties. In Section~\ref{sec:absorptionchars}, we describe the general characteristics of the sample of {\MgII} absorbers, discuss the degree to which it is a fair sample, and present selected kinematic properties of the absorbing systems. The VP fitting of these systems is described in Section~\ref{sec:vpanalysis} and the results of the VP fitting is presented and discussed in Section~\ref{sec:results}. For our work, we used the VP fitter {\sc Minfit} developed by \citet{churchill97}, which has since been upgraded and applied by  \citet{churchill03} and \citet{evans-phd}. We present our concluding remarks in Section~\ref{sec:conclude}.

\section{Data Analysis and Sample Building}
\label{sec:data}

\subsection{Spectra}
\label{subsec:spectra}

We have analyzed 249 high resolution, high signal-to-noise High Resolution Echelle Spectrometer \citep[HIRES,][]{vogt94} and Ultraviolet and Visual Echelle Spectrograph \citep[UVES,][]{dekker00} quasar spectra obtained from the Keck and Very Large Telescope (VLT) observatories, respectively.  The wavelength coverage of the spectra range from approximately 3,000--10,000~{\AA}, thought there is variable coverage in this range from spectrum to spectrum.

The resolving power of both instruments is $R = \lambda/\Delta\lambda = 45,000$, or $\sim 6.6$ {\kms}, and the spectra have three pixels per resolution element.  The resolution is approximately constant in velocity as a function of observed wavelength.  The signal-to-noise ratios are typically 25--80.  Such high quality spectra allow for an in-depth investigation into the distributions and kinematics of the galactic halos and intergalactic structures selected by the presence of metal line absorption.  

\begin{figure*}[htb]
\figurenum{1} \centering
\includegraphics[width=\textwidth]{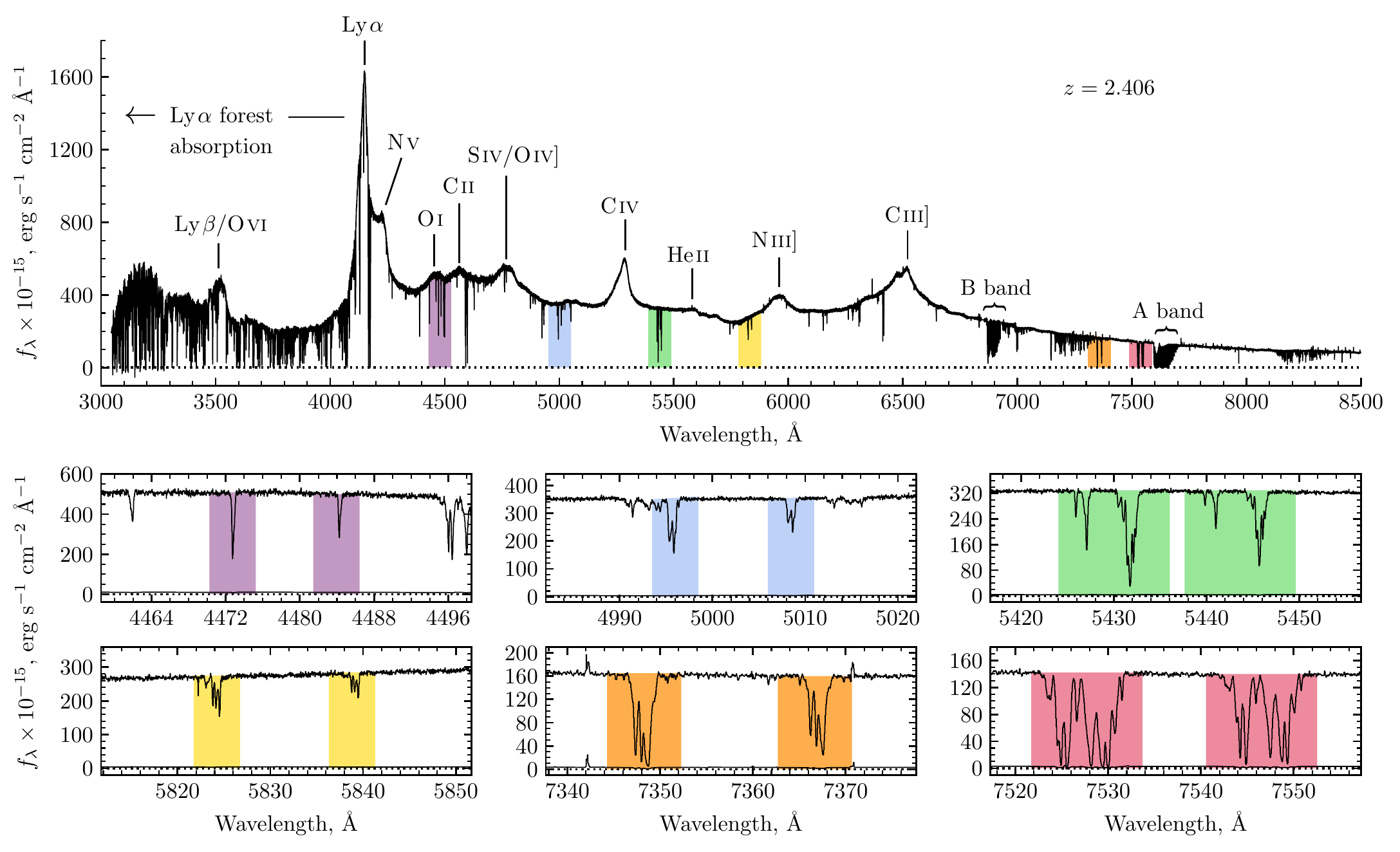}
\caption{The UVES spectrum J222006$-$280323 and six of its {\MgIIdblt} doublets presented in redshift order: $z=0.599508$ (magenta), $z=0.786512$ (blue), $z=0.942415$ (green), $z=1.082795$ (yellow), $z=1.555851$ (orange), and $z=1.692150$ (red).}
\label{fig:spectrum}
\end{figure*}

The Keck/HIRES spectra were obtained from various observing programs prior to the creation of the Keck Observatory Archive\footnote{\url{https://www2.keck.hawaii.edu/koa/public/koa.php}}, and were  donated to this work by Charles Steidel, J. Xavier Prochaska, Christopher Churchill, and by Michael Rauch and the late Wallace Sargent.   The spectra of Churchill and Steidel were explicitly observed for {\MgII} absorption \citep[e.g.,][]{churchill01, churchill03, steidel02} and were selected based upon previous studies which used low resolution spectra to detect {\MgII} absorption of equivalent widths $W_r(2796) \geq 0.3$ {\AA} \citep{sargent88, steidel92}.  Those of Sargent and Rauch were selected for high resolution analysis of {\Lya} forest and {\CIV} absorption, and those of Prochaska for damped Lyman alpha (DLA) absorption \citep{prochaska07}.  

The VLT/UVES spectra were acquired through the efforts of the UVES SQUAD prior to their Data Release 1 \citep{uves-squad}. These archived UVES spectra were obtained by several researchers for a variety of scientific purposes, including {\Lya} forest, Lyman-limit and damped {\Lya} systems, and {\MgII}, {\CIV}, {\OVI}, and other metal absorption-line studies. The heterogeneous selection bias of our sample is addressed in {\S} \ref{subsec:bias}.  

An example UVES spectrum, the $z_{\rm em}\!=\!2.406$ quasar  J222006$-$280323, is shown in Figure \ref{fig:spectrum}. Strong emission lines are labelled, as are the regions of {\Lya} forest absorption and A-band and B-band atmospheric absorption. The journal of quasar spectra used in this study is listed in Table~\ref{tab:journal}.  For each quasar, the columns list (1) the quasar name, taken from \citet{vcv01}, (2) a common B1950 name \citep[mostly from the catalog of][]{hb96}, (3) the emission redshift, (4) the lower wavelength limit observed, (5) the upper wavelength limit observed, and (6) the instrument with which the spectrum was obtained.

\begin{deluxetable*}{llrrrl}
\tablewidth{0pt}
\tablecaption{Journal of Observations\label{tab:journal}}
\tablehead{
\colhead{Quasar} & \colhead{Alias} &
\colhead{z$_{\rm em}$} & \colhead{$\lambda _{\rm blue}$} &
\colhead{$\lambda _{\rm red}$} & \colhead{Facility} \\[-4pt]
\colhead{\phantom{x}} & \colhead{\phantom{x}} & \colhead{\phantom{x}}
& \colhead{[\AA]} & \colhead{[\AA]} & \colhead{\phantom{x}}
}
\startdata
J000323$-$260318 & Q 0000$-$263 &      4.111 &       5122 &       8143 & HIRES\\[-4pt] 
J000149$-$015939 & UM 196 &      2.817 &       3045 &      10087 & HIRES, UVES\\[-4pt]
J000520$+$052411 & UM 18 &      1.900 &       3188 &       6081 & HIRES \\[-4pt]
J000344$-$232355 & HE 0001$-$2340 &      2.280 &       3044 &      10088 & UVES\\[-4pt] 
J000448$-$415728 & Q 0002$-$422 &      2.760 &       3044 &      10087 & UVES \\[-4pt]
$\cdots$ & $\cdots$ & $\cdots$ & $\cdots$ & $\cdots$ & $\cdot$ \\[-4pt]
$\cdots$ & $\cdots$ & $\cdots$ & $\cdots$ & $\cdots$ & $\cdot$ \\[-4pt]
$\cdots$ & $\cdots$ & $\cdots$ & $\cdots$ & $\cdots$ & $\cdot$ \\[-4pt]
J234625$+$124743 & B 234352$+$123103 &      2.578 &       3282 &      6651 & UVES \\[-4pt]
J234819$+$005721A & Q 23452$+$007A &      2.160 &       3285 &     6651 & UVES \\[-4pt]
J234825$+$002040 & BGCFH 46 &      2.650 &       3282 &      6550 & UVES \\[-4pt]
J235034$-$432559 & CTSC 15.05 &  2.885 &       3045 &      10086 & UVES \\[-4pt]
J235057$-$005209 & UM 184 &      3.024 &       3287 &      7493 & HIRES, UVES \\
\enddata
\tablecomments{Table~\ref{tab:journal} is published in its entirety in machine-readable format. A portion is shown here for guidance regarding its form and content.}
\end{deluxetable*}

\subsection{Reduction of Spectra}
\label{subsec:reduction}

The spectra of Churchill were reduced using the standard Image Reduction and Analysis Facility (IRAF\footnote{IRAF is distributed by the National Optical Astronomy Observatories, which are operated by the Association of Universities for Research in Astronomy, Inc., under cooperative agreement with the National Science Foundation.}) software, the process of which is detailed in \citet{churchill97}. Those of Sargent, Rauch, Prochaska, and Steidel were reduced using the Mauna Kea Echelle Extraction (MAKEE) data reduction package of \citet{barlow05}, which is optimized for the spectral extraction of single, unresolved point sources.  All HIRES spectra were wavelength calibrated using ThAr lamps to the vacuum heliocentric standard at rest and were continuum fit by their respective donors.

The UVES spectra were reduced using the UVES pipeline in the MIDAS environment \citep{dekker00}, which is provided by the European Southern Observatory (ESO).  The wavelength solution is determined using a standard ThAr lamp
exposure. Finally, the quasar flux is extracted and the ThAr wavelength calibration polynomial for each order is attached, having been corrected to vacuum heliocentric velocities. The indovodual exposures were then combined into one-dimensional spectra using the UVES POst-Pipeline Echelle Reduction (UVES {\sc Popler}) software \citep{murphy08, murphy16, uves-squad}, which is an extra reduction step that facilitates cosmic ray removal. An initial continuum is derived by iteratively fitting small sections of each spectrum with a low order Chebyshev polynomial, rejecting points lying many sigma below or above the fit at each iteration.  The sections of continuum are then spliced together by linearly weighting each section from unity at their centers to zero at their edges.  Any remaining artifacts, such as from internal spectrograph reflections, are removed and the continuum is refined manually.  

For 25 quasars, multiple spectra were available. We optimally combined them in order to exploit all wavelengths covered and to maximize signal-to-noise ratio.  Thus, in some cases, the final version of the spectrum we studied may comprise, for example, two unique HIRES observations and one UVES observation.  This provided a single spectrum for a quasar.  We ensured wavelength alignment by performing cross correlations on unresolved features in regions of overlapping wavelengths. Possible non-linear effects were accounted for by fitting a first-order polynomial to the cross correlation shifts as a function of wavelength. The flux in the pixels were optimally averaged, weighted by the inverse of their variances using flux conservation and pixel interpolation.

\subsection{Identifying Doublets and Systems}
\label{subsec:identifying}

We limited our search for absorption lines to wavelength regions spanning redward of the quasar {\Lya} emission line up to 5,000 {\kms} blueward of the quasar {\MgII} emission line. The lower wavelength limit ensures {\MgII} absorption is not confused with {\Lya} forest lines, and the upper limit is our adopted criterion for detected {\MgII} absorption features to not be considered associated with the quasar vicinity \citep[e.g.,][]{weymann91}. Using our software {\sc Search} \citep[e.g.,][]{weakI}, the spectra are objectively scanned for {\MgII} doublet candidates. The initial criteria for a candidate doublet are that the detection of a $5\sigma$ feature, which is taken to be the $\lambda$2796 line at redshift $z=\lambda/2796.352 - 1$, is accompanied by a corresponding feature with a $3\sigma$ detection at the projected location $\lambda = 2803.531\!\times\! (1+z)$. Doublet candidates, preliminary equivalent widths, and detection significance levels follow the formalism of \citet{schneider93}. The candidates are then checked for doublet ratios of $1.0 \leq DR \leq 2.0$ consistent within errors.  

Once a candidate {\MgII} doublet is identified, the {\sc Search} software is capable of locating and examining numerous associated atomic transitions simultaneously.  For this work, examination of the associated absorption features was limited to 13 commonly observed transitions from five abundant chemical elements.  Including the {\MgIIdblt} doublet, these transitions were {\MgI} $\lambda$2853, the {\FeII} $\lambda$2344, $\lambda$2374, $\lambda$2383, $\lambda$2587, and $\lambda$2600 quintuplet, the {\MnII} $\lambda$2577, $\lambda$2594, and $\lambda$2606 triplet, and the {\CaII} $\lambda\lambda 3935, 3970$ doublet.  The transitions and their adopted atomic data \citep{moore70, cashman17} are listed in Table \ref{tab:transitions}.  The columns are (1) the ion and transition, (2) the transition wavelength in vacuum, (3) the oscillator strength, and (4) the natural broadening, or damping constant.

\begin{deluxetable}{llll}
\tablewidth{0pt}
\tablecaption{Analyzed Transitions\label{tab:transitions}}
\tablehead{
\colhead{Ion / Transition} & \colhead{$\lambda_0$} &
\colhead{$f$} & \colhead{$\Gamma$ x $10^8$}\\[-4pt]
\colhead{\phantom{x}} & \colhead{[\AA]} & \colhead{\phantom{x}}
& \colhead{[sec$^{-1}$]}
}
\startdata
{\MgII}~$\lambda 2796$ & 2796.352 & 0.6123 & 2.612\\[-4pt]
{\MgII}~$\lambda 2803$ & 2803.531 & 0.3054 & 2.592\\[-4pt]
{\MgI} ~$\lambda 2852$  & 2852.964 & 1.8100 & 4.950\\[0.1cm]
{\FeII}~$\lambda 2344$ & 2344.214 & 0.1097 & 2.680\\[-4pt]
{\FeII}~$\lambda 2374$ & 2374.461 & 0.0282 & 2.990\\[-4pt]
{\FeII}~$\lambda 2382$ & 2382.765 & 0.3006 & 3.100\\[-4pt]
{\FeII}~$\lambda 2587$ & 2586.650 & 0.0646 & 2.720\\[-4pt]
{\FeII}~$\lambda 2600$ & 2600.173 & 0.2239 & 2.700\\[0.1cm]
{\MnII}~$\lambda 2576$ & 2576.877 & 0.3508 & 2.741\\[-4pt]
{\MnII}~$\lambda 2594$ & 2594.499 & 0.2710 & 2.685\\[-4pt]
{\MnII}~$\lambda 2606$ & 2606.462 & 0.1927 & 2.648\\[0.1cm]
{\CaII}~$\lambda 3934$ & 3934.777 & 0.6346 & 1.456\\[-4pt]
{\CaII}~$\lambda 3969$ & 3969.591 & 0.3145 & 1.414\\
\enddata 
\end{deluxetable}

Candidate {\MgII} doublet line profiles were aligned in rest-frame velocity and visually inspected to ensure that they exhibit velocity alignment and flux decrements consistent with the radiative transfer and atomic physics of the two transitions and to identify and annotate any blends or spurious spectroscopic features. For {\MgIIdblt} doublet features passing the criteria for inclusion into the sample, we adopt the definition that two or more {\MgII}~$\lambda 2796$ absorption features comprise a single {\it absorption system\/} if they reside within $\pm 800$~{\kms} of each other\footnote{Adopting a velocity ``window'' of $\pm 410$~{\kms} to $\pm 1030$~{\kms} yields an identical sample of systems and, subsequently, identical distributions in the system properties.}. The corresponding transitions listed in Table~\ref{tab:transitions} are also visually checked. 

\begin{deluxetable*}{lrrlrr}
\tablewidth{0pt}
\tablecolumns{6}
\tablecaption{System Properties\label{tab:samplesysanaldata}}
\tablehead{
\colhead{Quasar} & \colhead{$z_{\rm abs}$} &
\colhead{$W_{r}(2796)$} & \colhead{$\log\! N_{\rm aod}$} &
\colhead{$DR$} & \colhead{$\omega_{v}$} \\[-4pt]
\colhead{\phantom{x}} & \colhead{\phantom{x}} & \colhead{[\AA]} &
\colhead{[cm$^{-2}$]} & \colhead{\phantom{x}} & \colhead{[km~s$^{-1}$]}
}
\startdata
$\mathrm{J012417\!-\!374423}$ &   1.173635 & $     0.018\pm     0.001$ & $     11.63_{-      0.07}^{+      0.09}$ & $      1.84\pm      0.30$ & $       6.4\pm       1.0$\\[0.03cm] 
$\mathrm{J101447\!+\!430031}$ &   2.042606 & $     0.092\pm     0.004$ & $     12.39_{-      0.03}^{+      0.05}$ & $      1.85\pm      0.17$ & $      59.6\pm       1.6$\\[0.03cm] 
$\mathrm{J123200\!-\!022404}$ &   0.756903 & $     0.303\pm     0.003$ & $     13.30_{-      0.06}^{+      0.07}$ & $      1.28\pm      0.02$ & $      15.6\pm       0.7$\\[0.03cm] 
$\mathrm{J110325\!-\!264515}$ &   1.202831 & $     0.593\pm     0.002$ & $     13.44_{-      0.04}^{+      0.04}$ & $      1.44\pm      0.01$ & $      43.3\pm       0.2$\\[0.03cm] 
$\mathrm{J110325\!-\!264515}$ &   1.838689 & $     1.044\pm     0.001$ & $     13.80_{-      0.03}^{+      0.04}$ & $      1.33\pm      0.01$ & $      50.0\pm       0.1$\\[0.03cm] 
$\mathrm{J035405\!-\!272421}$ &   1.405188 & $     2.660\pm     0.006$ & $\geq     14.35$ & $      1.10\pm      0.01$ & $      93.3\pm       0.3$\\
\enddata
\tablecomments{Table~\ref{tab:samplesysanaldata} is published in its entirety in machine-readable format. A portion is shown here for guidance regarding its form and content.}
\end{deluxetable*}

Within the higher wavelength regimes of our sample (above $\sim 6800$ {\AA}), corresponding to {\MgII} $\lambda$2796 absorption redshifts of $z \geq 1.43$, feature identification becomes more difficult due to the presence of telluric lines.  The strongest lines occur in the A- and B-bands (7600--7630 and 6860--6890 {\AA}, respectively) and also between 7170--7350 {\AA} \citep[e.g.,][]{barlow05}. These line complexes are generally distinguishable from the {\MgIIdblt} doublet.  Except within the highly saturated bands themselves, the A- and B-band lines exhibit distinctive patterns of closely spaced pairs. A chance alignment of two A- or two B-band telluric lines at the precise separation of a candidate {\MgII} doublet, along with the required doublet ratio, is not common. Nevertheless, these spurious features may obscure weak {\MgII} absorption lines or cause some confusion due to line blending. In addition to our objective criteria for identifying {\MgII} doublet candidates using the {\sc Search} algorithm, we were especially diligent to visually inspect candidates in these wavelength regions. Corroboration by visual inspection of the associated  {\MgI}, {\FeII}, {\CaII}, and {\MnII} features mentioned in {\S}~\ref{subsec:identifying} was also also performed, but a corroboration was not required for the candidate to be included in the sample. These procedures were adopted in order to maximize the accuracy in the number of absorbing systems located in the spectra.

Using these criteria, we found a total of 480 {\MgII} absorption systems, covering the redshift $0.19 \leq z \leq 2.55$ and having rest-frame equivalent widths over the range $0.006 \leq W_r(2796) \leq 6.23$~{\AA}.  These 480 systems were found in 186 of the 249 quasars we searched.

\subsection{Measuring Doublets and Systems}
\label{subsec:measuringing}

Once a {\MgII} system was confirmed, we used our code {\sc Sysanal} \citep[see][]{weakI, churchill01} to analyze the absorption. The code first computed the optical-depth median of {\MgII}~$\lambda 2796$ profile, i.e., the wavelength ($\lambda_{\bar{\tau}}$) at which equal integrated optical depth resides to both sides of the profile. By definition, we adopt the system redshift as $z_{\rm abs}= \lambda_{\bar{\tau}}/2796.352 - 1$, which is employed to compute the systemic rest-frame velocity zero point of the absorption system.  The code then computed the rest-frame equivalent widths, $W_r$(2796), the {\MgII} doublet ratios, $DR$, the apparent optical depth column densities, $N_{\rm aod}$, the kinematic velocity spreads, $\omega_v$, and errors in these quantities for all transitions.  The apparent optical depth column density was obtained by inverting the absorption profile to an optical depth profile using the radiative transfer solution $\tau_\lambda = \log (I^0_\lambda / I_\lambda)$, where $I^0_\lambda$ is the fitted continuum, converting the optical depth to column density per unit velocity, and integrating over the profile \citep{savage91},
\begin{equation}
N_{\rm aod} = \frac{m_e c^2}{\pi e^2} \frac{1}{f\lambda_0^2} 
\int \!\! \tau_\lambda \, d\lambda \, .
\end{equation}
Lower limits on $\tau_\lambda$ occur when the pixel at $\lambda$ and its two adjacent pixels meet the condition $I_\lambda < \sigma_{I_\lambda}$, in which case $\tau_\lambda \geq \log (I^0_\lambda / \sigma_{I_\lambda})$.  If this condition occurs in three contiguous pixels, corresponding to a resolution element, then the profile is considered to be saturated and $N_{\rm aod}$ is quoted as a lower limit for the apparent optical depth column density.

The kinematic velocity spread is the proportional to the flux-decrement weighted second-moment of the velocity across the {\MgII}~$\lambda 2796$ absorption profile \citep{sembach92},
\begin{equation}
   \omega^2_v =  \frac{V_{(2)}}{V_{(0)}} = \frac{1}{V_{(0)}} \int \!\! \left( v - \langle v \rangle \right) ^2 D(v)\, dv \, ,
\end{equation}
where $D(v) = 1 - I(v)/I^0(v)$ is the flux decrement in velocity coordinates, $\langle v \rangle$ is the mean velocity (the flux-decrement weighted first-moment of the velocity), and $V_{(0)}$ is the velocity ``equivalent width" (zeroth moment), 
\begin{equation}
  \langle v \rangle = \int \!\! v D(v)\, dv \, , \qquad  V_{(0)} = \int \!\! D(v)\, dv \, .
\label{eq:meanv}
\end{equation}
The kinematic velocity spread, $\omega_v$, can be interpreted as the equivalent Gaussian standard deviation of the absorption profile. For each associated transition, if absorption is not formally detected at the $3\sigma$ significance level at the expected location in the spectrum, the {\sc Sysanal} code computes the $3\sigma$ upper limits on the equivalent widths and apparent optical depth column densities,

\section{Absorption Characteristics}
\label{sec:absorptionchars}



In Table \ref{tab:samplesysanaldata}, we present the measured properties of the 480 {\MgII} absorption systems we analyzed using the {\sc Sysanal} code. Tabulated are (Column 1) the quasar name, (2) the system redshift, (3) the {\MgII} $\lambda$2796 rest-frame equivalent width, (4) the {\MgII} apparent optical depth column density, (5) the {\MgIIdblt} doublet ratio, and (6) the kinematic velocity spread.  Table~\ref{tab:samplesysanaldata} is published in its entirety in machine-readable format. A portion is shown here for guidance regarding its form and content.

\begin{figure}[bht]
\figurenum{2} \epsscale{1.0} 
\plotone{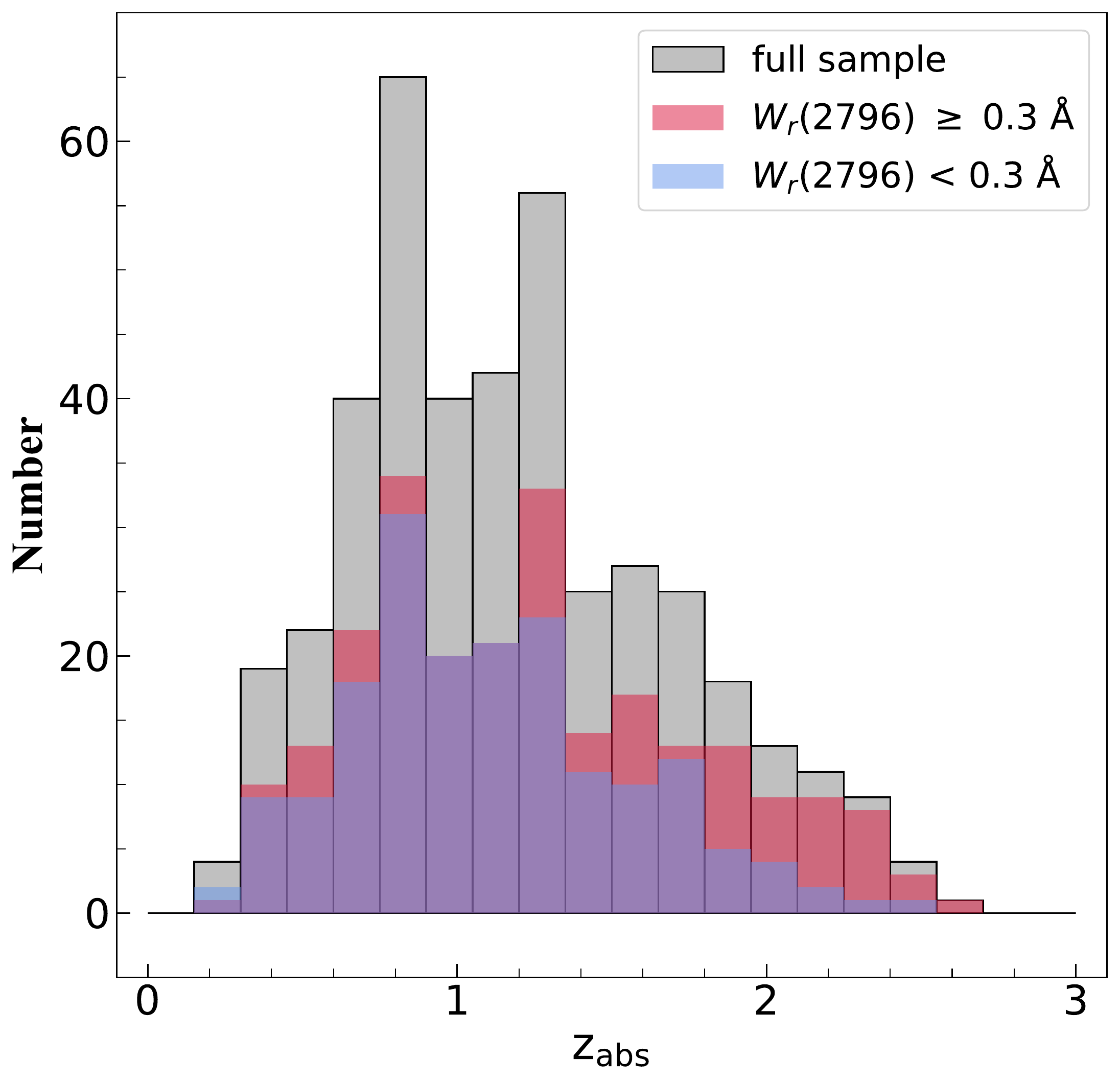}
\caption{ The distribution of {\MgII} absorption redshifts for the complete sample of 480 systems identified in our survey. The solid gray area indicates the full sample.  The pink area shows the strong systems and the blue area shows the weak systems. Superposition of the weak and strong systems appears purple.}
\label{fig:zabs}
\vglue -0.1in
\end{figure}

In Figure~\ref{fig:zabs}, we present the redshift distributions of the 480 systems. The full sample has is shown in gray. The absorber redshifts cover the range $0.19\! \leq\! z \leq 2.55$, with $\langle z \rangle \!=\! 1.18$, and comprise systems with $0.006 \leq W_r(2796)\! \leq\! 6.23$ {\AA}.  The overall shape of this distribution is governed primarily by the summed redshift path coverage of the quasar spectra, which differs in each redshift bin. This redshift path coverage diminishes toward lower redshift and toward higher redshift.  As we will not be examining the evolution of absorption properties in this paper, we report only the distribution of the observed sample, and do not discuss the {\MgII} redshift path coverage nor the redshift path coverage sensitivity functions \citep[e.g.][]{lanzetta87, steidel92, weakI, nestor05}.

Historically, {\MgII} absorbers were divided into ``weak" systems \citep[defined to have $W_r(2796) < 0.3$~{\AA},][]{weakI, rigby02} and ``strong" systems \citep[defined to have $W_r(2796) \geq 0.3$~{\AA},][]{steidel92}. This was due to the equivalent width detection sensitivity of $\simeq 0.3$~{\AA} of 3-meter class telescopes and lower-resolution spectrographs of the late 1980s and early 1990s, which was significantly reduced by an order of magnitude with the advent of the Keck 10-meter telescope and the HIRES spectrograph \citep{vogt94}. We adopt this historical definition. 

In Figure~\ref{fig:zabs}, the redshift distribution of the weak systems is shown in blue whereas the distribution for strong systems is shown in pink. The superposition of these two populations appears purple.  The more sharply declining tail at the higher redshift region of the weak population versus that of the strong population is due partially to a decline in the equivalent width detection threshold of the spectra in the higher redshift regime; the lowest $W_r$ systems suffer the most loss of detection completeness due to the increase in telluric lines. The other reason there is a more rapid decline in the weak absorbers at higher redshift is that the number density of such absorbers per unit redshift decreases rapidly beyond $z\sim 1.5$ \citep[see][and references therein]{weak12}.

\begin{figure*}[htb]
\figurenum{3} \centering
\includegraphics[width=0.8\textwidth]{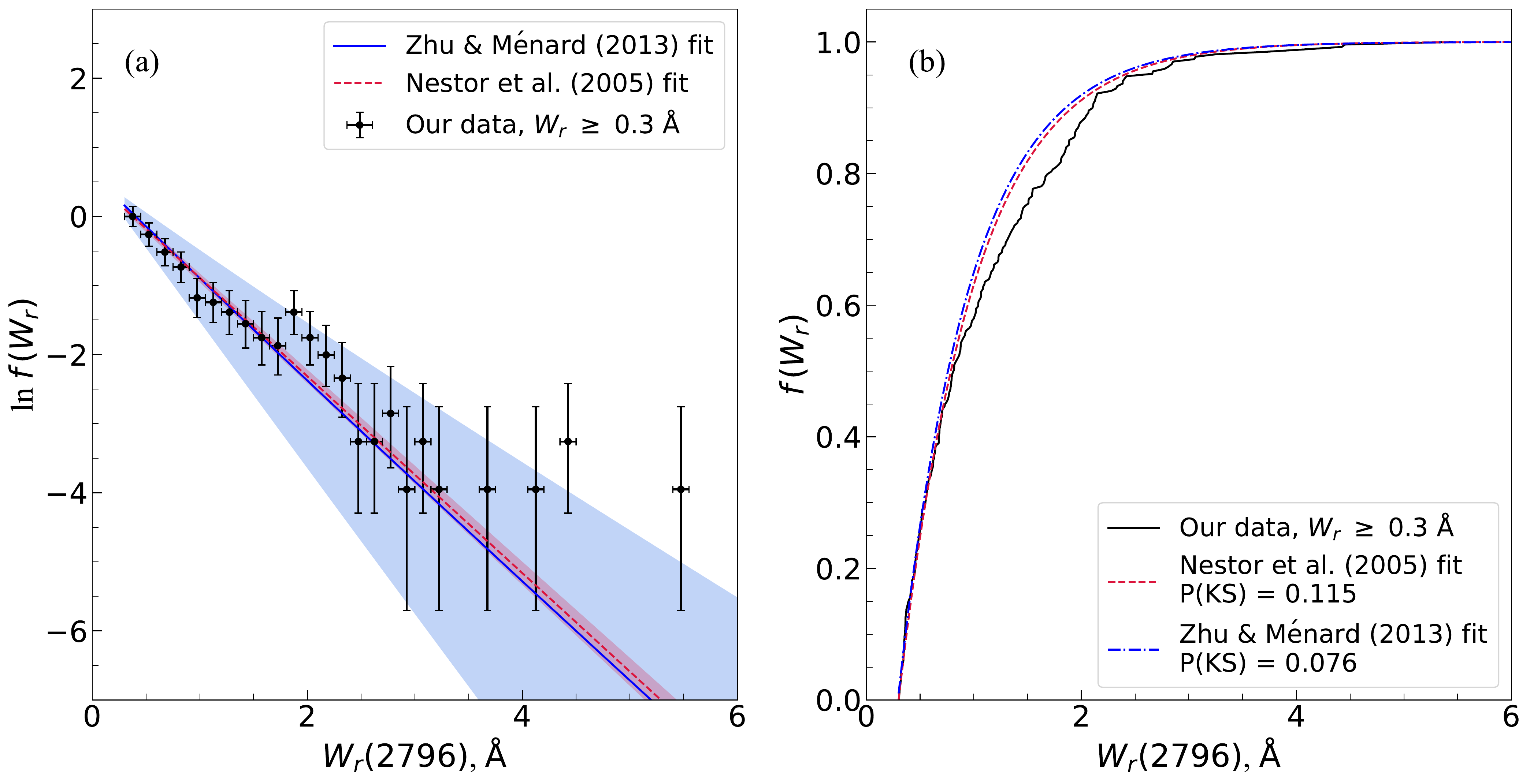}
\caption{ (a) The binned {\MgII} $\lambda 2796$ equivalent width distribution of our complete for $W_r(2796) \geq 0.3$ {\AA} limited to the redshift range $0.34 \leq z < 2.27$, which included 469 of the complete sample of 480 systems. The green shaded region superimposes the slope and uncertainty in the maximum likelihood exponential fit of \citet{nestor05}. The purple shaded region superimposed the parameterized fitted model and uncertainty from \citet{zhu13}. The Nestor and Zhu \& M\'enard parameterized functions are normalized to the area under the observed distribution. (b) The cumulative distribution of our sample compared to the cumulative distributions of the Nestor and Zhu \& M\'enard models.  K-S tests yielded $P({\rm KS})=0.115$ and $P({\rm KS})=0.076$ indicating that the observed equivalent width distribution is not inconsistent with the unbiased distributions. Thus, we expect the general distribution of kinematic properties of our sample to also reflect that of an unbiased absorption path normalized sample.}
\label{fig:ntrewdist}
\end{figure*}

\begin{figure*}[hbt]
\figurenum{4} \centering
\includegraphics[width=0.8\textwidth]{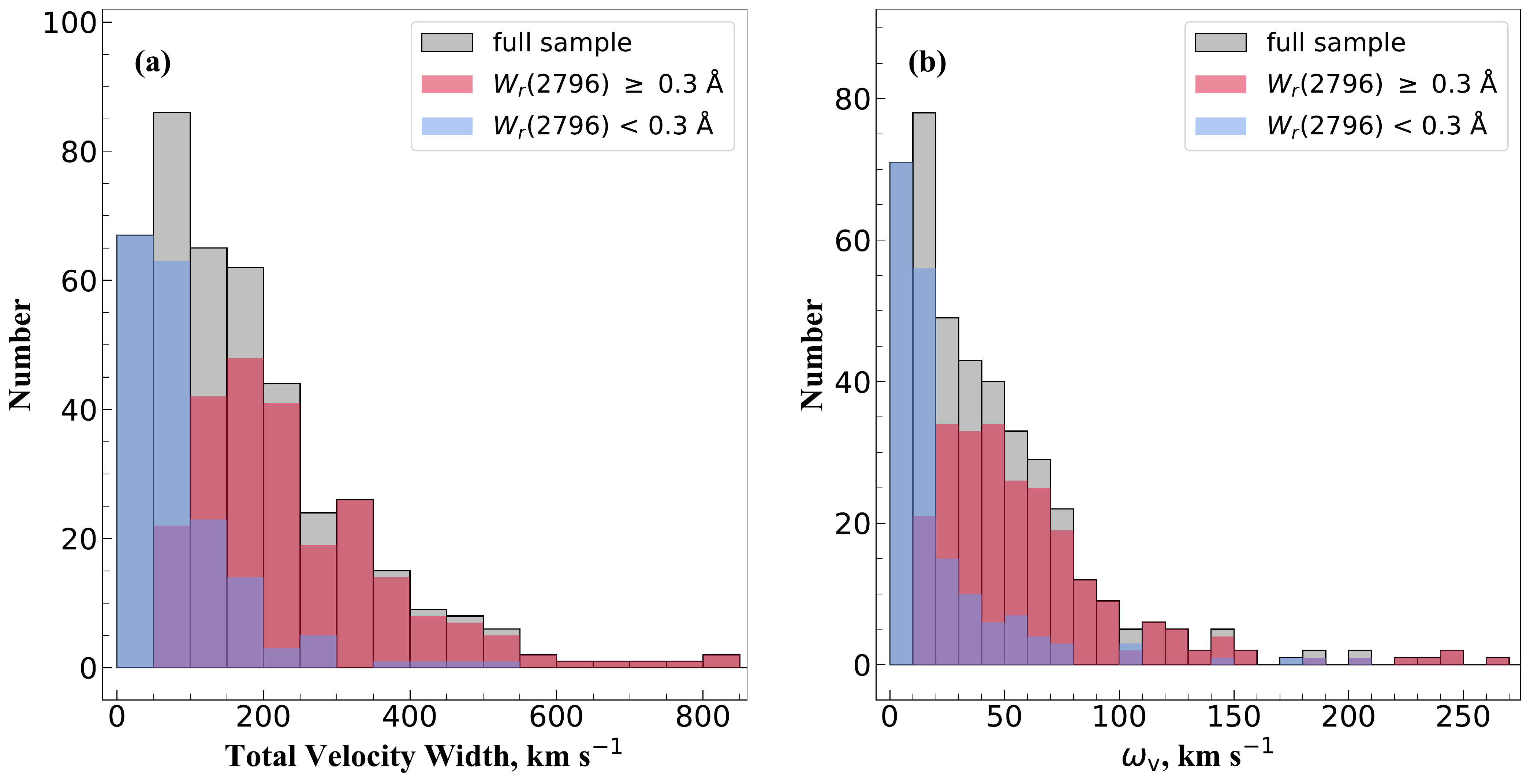}
\caption{(a) The distributions of {\MgII}~$\lambda 2796$ total velocity widths for the 422 systems in the kinematic sample, showing all systems (grey), weak systems (blue), and strong systems (pink); superposition of the weak and strong systems appears purple.  (b) The distributions of the kinematic velocity spread, $\omega_v$, using the same color scheme as panel (a).}
\label{fig:veldists}
\end{figure*}

\subsection{Examination of Sample Bias}
\label{subsec:bias}

As previously mentioned, we do not examine redshift evolution of the absorption properties in this paper. However, as we do aim to present and discuss the distribution of the kinematic properties over the redshift range we surveyed, it is important we establish that we have a reasonably fair sample of {\MgII}-selected absorption systems. 

It is well-established that {\MgII}~$\lambda 2796$ rest-frame equivalent width, $W_r(2796)$, and various kinematic indicators are correlated. For example, the kinematic velocity spread, $\omega_v$ is positively correlated with $W_r(2796)$ \citep[e.g.][]{archiveI, churchill01}, as is the number of Voigt profile components \citep[e.g.,][]{petitjean90, churchill03}. Thus, we adopt the premise that a sample of {\MgII}-selected absorbers with a fair distribution of {\MgII}~$\lambda 2796$ rest-frame equivalent widths would also represent a fair distribution of {\MgII} kinematic properties.

Since none of our quasar spectra were selected based upon knowledge of weak absorption, and because no correlation exists between strong and weak absorbers in a given quasar spectrum \citep{weakI}, our sample is
unbiased toward weak systems.  However, some of the quasar spectra were observed because the presence of strong {\MgII} absorption had already been ascertained from previous low-resolution surveys.  Given this partial selection bias and the heterogeneous scientific motivations behind the observations of many of these lines of sight (as discussed in Section~\ref{subsec:spectra}), we cannot {\it a priori} expect that the strong {\MgII} absorption subset is consistent with an unbiased sample.  Some statistically founded reassurance of this would allow us to adopt the view that our sample is a fair sample for studying the kinematic aspects of the strong {\MgII} absorption systems.

In order to determine whether our strong sample is fair and unbiased, we quantitatively compare the distribution of equivalent widths for $W_r(2796 \geq 0.3$~{\AA} from our sample to those from the large blind SDSS surveys of \citet[][1331 {\MgII} absorbers]{nestor05} and of \citet[][40,000 {\MgII} absorbers]{zhu13}. To allow direct comparisons between all three surveys, we limited our analysis to the redshift range of \citet{nestor05}, which is in common with our survey and that of \citet{zhu13}. This redshift range is $0.34 \leq z_{\rm abs} \leq 2.27$, comprising 469 out of the total of 480 absorbers in our survey.  For the \citet{nestor05} distribution function, we adopted their exponential fit to the function $f(W) = (N^*/W^*)\exp \{-W/W^*\}$, where  $N^* = 1.187 \pm 0.052$ and $W^* = 0.702 \pm 0.017$~{\AA}, which applies for $W_r(2796 \geq 0.3$~{\AA}.  For the \citet{zhu13} distribution function, we adopted their Eq.~5 for $dN^2/dWdz$ and best fit parameters from their Table~2. We integrated this function over the adopted redshift range to obtain $f(W) = dN/dW$. We then normalized both distribution to the area under the distribution for the observed data from our survey.


In Figure~\ref{fig:ntrewdist}(a), we present the binned {\MgII}~$\lambda 2796$ rest-frame equivalent width distribution normalized to unity at $W_r(2796)=0.3$~{\AA}. The normalized \citet{nestor05} and \citet{zhu13} distribution functions are superimposed on the data. The shaded regions account for the uncertainties in the fit parameters. Visual inspection would suggest that our sample of strong {\MgII} absorbers is populated by an slight overabundance of absorbers having $W_r(2796) \sim 2$~{\AA} and perhaps also having $W_r(2796) > 4$~{\AA}, but that otherwise, within measurement uncertainties, our distribution has a shape generally consistent with an exponential distribution.
 

In Figure~\ref{fig:ntrewdist}(b), we plot the cumulative equivalent width distributions. For the observed data, the relative decrement in the range $1 \leq W_r(2796) \leq 2$~{\AA} reflects our slight overabundance of $W_r(2796) \sim 2$~{\AA} equivalent width systems as compared the unbiased surveys. A Kolmogorov-Smirnov (K-S) statistical test was employed in order to quantitatively measure the consistency (or lack thereof) between the distribution of our sample of strong {\MgII} systems and the unbiased distributions.  We adopt the criterion that a probability for a K-S statistic indicative of the two distributions being inconsistent with one another is $P({\rm KS}) \leq 0.0027$, corresponding to a $3\sigma$ significance level. Compared to the \citet{nestor05} distribution, we obtained $P({\rm KS}) = 0.115$. Compared to the \citet{zhu13} distribution, we obtained $P({\rm KS}) = 0.076$.  Thus, both tests indicate the observed distribution cannot be ruled inconsistent with the unbiased surveys. Even through the observed cumulative distributions exhibit some shape variation compared to the unbiased surveys, this variation is significant only at the $1.6\sigma$ and $1.8\sigma$ levels for the \citep{nestor05} and \citep{zhu13} distributions, respectively.

The minor discrepancy is likely due to some of our quasar lines of sight being observed because a strong absorber, such as a DLA, was targeted for high-resolution analysis. For example, the HIRES spectra contributed by J.~X. Prochaska, as well as some of the UVES spectra from the VLT archive \citep[e.g.,][]{jorgenson13}, were obtained for DLA studies, which are known to exhibit {\MgII} absorption with higher equivalent widths \citep[e.g.,][]{rao00}.  The excess at $W_r(2796) \geq 4$~{\AA} may also be indicative of lines of sight targeted for very large $W_r(2796)$ systems, perhaps for galactic wind studies \citep[e.g,][]{bond01a,bond01b,mas-ribas18}. As a result of this analysis, we proceed under the assumption that our sample of {\MgII} absorption systems has the characteristics of an unbiased sample for the purpose of examining general kinematic properties.

\subsection{Kinematic Properties}
\label{subsec:kinematic}

Studies of the kinematics of the absorption systems must account for variations in the signal-to-noise ratio across the velocity window over which the analysis is performed. A varying ratio can result in variable detection sensitivities for very weak components at various velocities locations across a {\MgII} $\lambda 2796$ profile; this could introduce systematics into the distribution of fitted VP component velocities.  
It is thus imperative we have a controlled sample for our VP and kinematic analysis; we need a uniform detection sensitivity to ensure weak, blended, and high-velocity components above a fixed equivalent width threshold are detectable in all systems included in the analysis.  Here, we describe our selection of this final science subsample (comprising 422 systems), which we use for the VP fitting, and analysis of the column density, $b$ parameter, and kinematic distributions of the VP components.

We created a ``kinematic sample'' by including only those systems for which a $5\sigma$ minimum rest-frame equivalent width detection limit was present over a velocity window of $\pm600$~{\kms} relative to the velocity zero-point of the {\MgII} $\lambda 2796$ absorption feature. The criterion for a system to be included in the kinematic sample was that either (1) the average limit is $\langle W_r^{\rm lim} \rangle \leq 0.02$~{\AA}; or (2) if $\langle W_r^{\rm lim} \rangle$ was greater than 0.02~{\AA}, then $\langle W_r^{\rm lim} \rangle$ minus the standard deviation in $\langle W_r^{\rm lim} \rangle$ was less than or equal to 0.02~{\AA}.  A total of 422 systems, or 88\% our complete sample of 480, met the criterion for inclusion into the kinematic sample. 

In Figure~\ref{fig:veldists}(a), we show the distribution of the system total velocity width for the kinematic sample.  The system total velocity width is defined as $\Delta v_{\rm tot} = \left| v_{\rm max} \!-\! v_{\rm min} \right|$, the absolute difference of the ``reddest" absorbing pixel and the ``bluest" absorbing pixel.  The color scheme for the binned date are the same as for Figure~\ref{fig:zabs}.  

The two populations exhibit markedly different distributions.  The distribution of $\Delta v_{\rm tot}$ for the weak systems can be modeled as a half-Gaussian centered on $\Delta v_{\rm tot} = 0$~{\kms} with a standard deviation of $\sigma (\Delta v_{\rm tot}) \simeq 90$~{\kms}. Weak systems with $\Delta v_{\rm tot} \geq 300$~{\kms} are very rare; these systems would comprise several weak components highly separated in velocity (see the $z_{\rm abs} = 2.042606$ systems along the J101447$+$430031 line of sight in Figure~\ref{fig:profile1}, which has $\Delta v_{\rm tot} \simeq 150$~{\kms}). The distribution for the strong systems can be modeled as an asymmetric Gaussian with a mode at $\Delta v_{\rm tot} \simeq 200$~{\kms} and an average standard deviation of $\sigma (\Delta v_{\rm tot}) \simeq 200$~{\kms}. The long tail extends out to $\Delta v_{\rm tot} \simeq 800$~{\kms} due to a small contribution of kinematically extreme systems. In our sample, there are no strong systems with $\Delta v_{\rm tot} \leq 50$~{\kms}. 

The total velocity spread measures the extremes of the absorption velocities; it contains no information about the flux decrement distribution. In Figure~\ref{fig:veldists}(b), we show the distribution of the kinematic velocity spread, $\omega_v$, also using the same color scheme as Figure~\ref{fig:zabs}.  Though the kinematic velocity spread contains information about the velocity distribution of the flux decrements, the weak and strong systems exhibit similar distribution shapes as for $\Delta v_{\rm tot}$. 

The distribution of $\omega_v$ for the weak systems can be modeled as a half-Gaussian centered on $\omega _v = 0$~{\kms} with a standard deviation of $\sigma (\omega_v) \simeq 15$~{\kms}. Weak systems with $\omega_v \geq 80$~{\kms} are very rare. The distribution for the strong systems can be modeled as an asymmetric Gaussian with a mode at $\omega_v \simeq 35$~{\kms} and a standard deviation of $\sigma (\omega_v) \simeq 50$~{\kms}. The long tail extends out to $\omega_v \simeq 250$~{\kms}. Note that there are no strong systems with $\omega _v \leq 10$~{\kms}.

\section{Voigt Profile Fitting}
\label{sec:vpanalysis}

One of our goals is to characterize ``cloud" kinematics, column density, $N$, and Doppler $b$ parameter distributions for the {\MgIIdblt} transitions. We thus have applied Voigt profile (VP) decomposition (or fitting) to the kinematic sample of 422 absorbers. The Voigt function models the optical depth of the absorption profiles while incorporating the atomic physics of the transitions, including the transition wavelengths, $\lambda_0$, the oscillator strengths, $f$, and the natural line broadening via the damping constants, $\Gamma$. The natural line broadening function is a Lorentzian centered on the transition wavelength with a half-height half-width of $\Gamma \lambda^2_0/4\pi c$ and amplitude proportional to $N\!f$. The Voigt function also incorporates additional line broadening via convolution of a Gaussian wavelength redistribution function.  This property makes the Voigt function ideal for modeling absorption lines in warm/hot gas, as the line-of-sight projected thermal distribution of atomic motions in an isothermal gas is a Gaussian function of Doppler width $\Delta \lambda_{\hbox{\tiny D}} = \lambda_0 (b/c)$, where $b=\sqrt{2kT/m}$ is the thermal Doppler $b$ parameter. 

Once the optical depth, $\tau_\lambda$, of an absorption line is modeled, one accounts for the solution to the equation of radiative transfer, i.e., $I_\lambda = I^0_\lambda \exp \{ -\tau_\lambda \}$ to obtain the observed counts across the profile. Finally, to account for the instrumental line spread function, $\Phi (\Delta \lambda)$, the model of $I_\lambda$ is convolved with $\Phi (\Delta \lambda)$. As $N$, $b$, and 
\begin{equation}
v = c \, \frac{\lambda - \lambda_0(1+z)}{\lambda_0(1+z)} \, ,
\end{equation}
are adjustable parameters of the optical depth model, where $v$, $\lambda$, and $z$ are the line center rest-frame velocity, observed wavelength, and redshift, respectively, one can employ least squares fitting, most commonly using the $\chi^2$ statistic, to obtain estimates and uncertainties in $N$, $b$, and $z$ (the latter being converted to $v$).

Assuming that complex absorption profiles comprise multiple isothermal ``clouds'', each with a unique line-of-sight velocity, we can decompose absorption line systems into multiple Voigt profiles, with each model component yielding a column density, Doppler $b$ parameter, and line-of-sight rest-frame velocity for the ``cloud" being represented.

\subsection{Fitting Approach}
\label{subsec:phil}

Central to our fitting approach is that we adopt a minimalist approach to the modeling. We fit the absorption systems using as few ``clouds", or VP components, as possible by ensuring all components are statistically significant to the $\chi^2$ statistic.  Details of how we ensure statistical significance of all components are given in Section~\ref{subsec:procedure}.

The ionization potentials of the five ions included in the fit are all within a few to several electron volts (eV) of the {\HI} potential of 13.59 eV. The {\MgI} and {\CaII} ionization potentials are both below that of {\HI} at 7.65 eV and 11.87 eV, respectively, while those of {\MgII}, {\FeII}, and {\MnII} are slightly above at 15.04 eV, 16.19 eV, and 15.63 eV, respectively.  This bracketing of the {\HI} ionization potential does mean that the line-of-sight ionization structure of a given parcel of gas could be different for the different ions. We assume that any difference is negligible within the context of the resolution and signal-to-noise ratio of the spectra and the general VP decomposition premise of spatially separated isothermal ``clouds''.  The  absorption is therefore assumed, for the purpose of VP fitting, to arise all in the same spatial extent of gas and to reflect the same velocity structure.  

Another consideration in VP decomposition is how to fit the Doppler $b$ parameter across ions.  For our study, we have chosen to let the user input whether to constrain this parameter as 100\% ``turbulent" or 100\% thermal.  In the former case, the $b$ parameter of a given ``cloud", or VP component, is enforced to be identical for all ions, even though the $b$ parameter is still fit as a free parameter for each component. In the latter case, the $b$ parameter in a given component is enforced to scale in proportion to the inverse of the square root of the atomic mass for each ion.  We adopted the default condition of ``turbulent" broadening; departures from this are noted in the descriptions of individual systems in Appendix~\ref{sec:appendix}.  The systems were fit under the thermal condition if a satisfactory fit could not be achieved using the ``turbulent" condition.

In addition, VP modeling by nature requires the assumptions of isothermal ``clouds" that occupy distinct line-of-sight locations in velocity space.  These latter two assumptions are probably the least defensible given our developing understanding of the complexity of gas properties as gleaned from ``mock absorption line" studies of hydrodynamic cosmological simulations \citep[e.g.,][]{cwc-direct, peeples19}. However, \citet{cwc-direct} did find that in Eulerian adaptive mesh simulations, for low-ionization ions such as {\MgII}, the {\it absorbing\/} gas is cloud-like in that gas cells selected by detected absorption are spatially contiguous and comprise a narrow temperature range. \\

\subsection{Fitting Procedure}
\label{subsec:procedure}

For each absorption system, the first step is to create an initial model of the VP components.  We use our graphical interactive program {\sc iVPfit}, which is an improved version of {\sc Profit} \citep{churchill97}.  The output of {\sc iVPfit} is a complete model of all components for all transitions for all ions included in an absorption system.  The component parameters for all transitions of a given ion are ``tied", meaning that for a given ``cloud" the $N$, $b$, and $v$ for that ion are simultaneously constrained by all transitions of that ion and they all have the same value. The interactive process is streamlined by the use of auto-scaling of column densities between ions. We found, in the course of this study, that our final models were not sensitive to our ``$\chi$ by eye'' methodology, in which the user tended to be biased toward introducing more components than are statistically significant. For example, we never experienced a case where adding a greater number of components to an initial model changed the final number of components determined to be significant by {\sc Minfit}.

\begin{figure*}[htb]
\figurenum{5} \epsscale{1.10}
\plotone{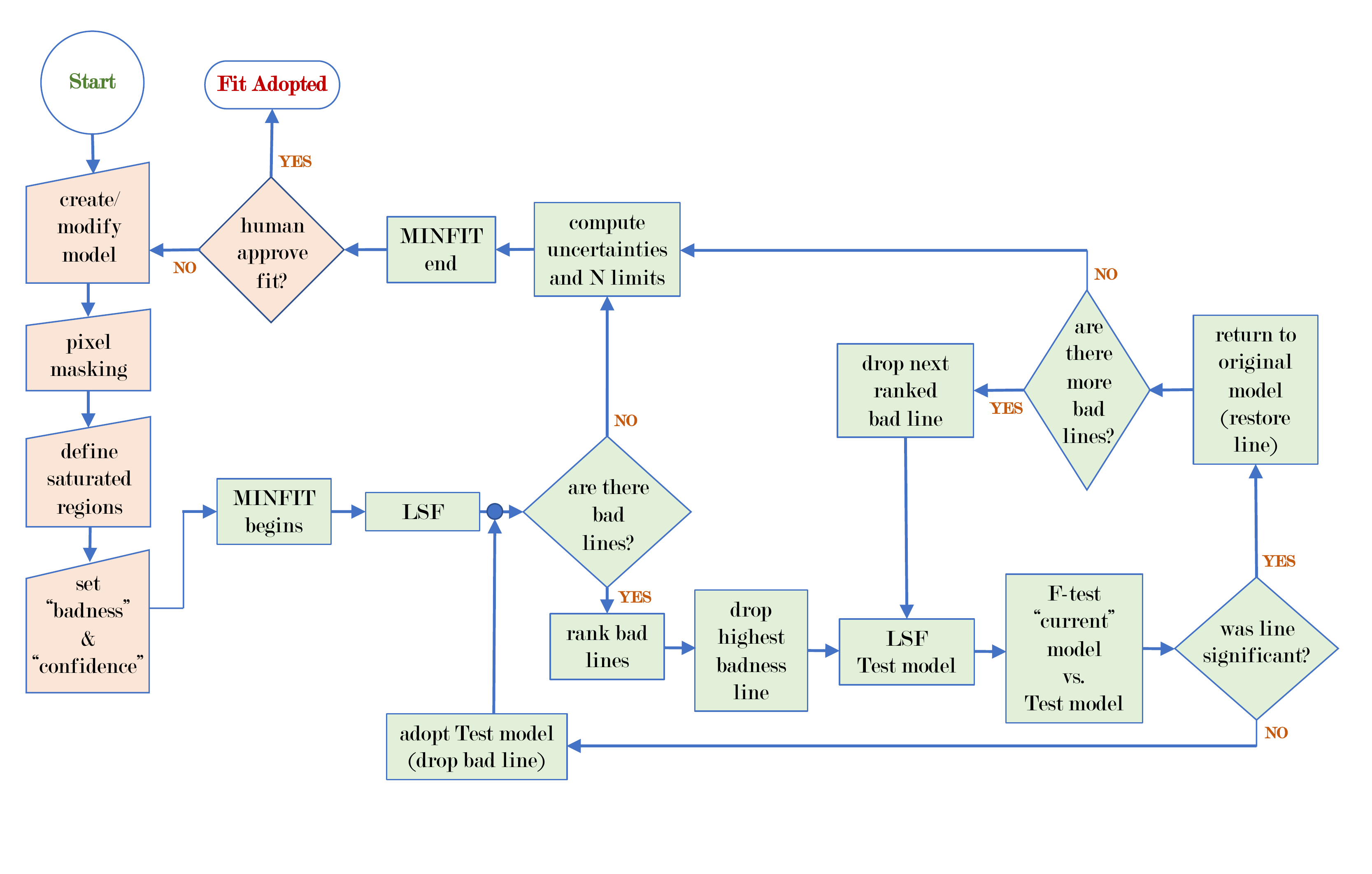}
\vglue -0.4in
\caption{The Voigt profile modeling process. Human involvement (peach boxes) occurs in the creation of an initial model, pixel masking, defining saturated regions, the definition of the badness and confidence parameters, and then again following the fully automated {\sc Minfit} decision tree (light green boxes), which begins with the least squares fitting (``LSF") and ends with the computation of uncertainties in measured VP parameters and column density limits for non-detections. The visual judgement of the goodness of the fit is the final human involvement.}
\label{fig:flowchart}
\vglue 0.1in
\end{figure*}

Using this initial model, we then employ the least squares fitter {\sc Minfit}\footnote{\url{https://github.com/CGM-World/minfit}.} \citep{churchill97}, which iteratively eliminates all statistically insignificant components and adjusts the remaining components until the least squares fit is achieved. This entire process, discussed below and illustrated in Figure~\ref{fig:flowchart}, is based on a series of objective tests and trials. {\sc Minfit} utilizes the spectral information from all available transitions to constrain these parameters.  The velocity of a given component is constrained to be the same across all transitions of all ions.  The column density is constrained to be the same across all transitions of a given ion.  Finally, the Doppler $b$ parameter can be set by the user to either vary thermally across ions or to be constant across all ions (the ``turbulent" option, as discussed in Section~\ref{subsec:phil}).

In a given system, one or more transitions may be compromised over portions of its velocity extent by either bad/noisy pixels or spurious features which have no corroborating absorption in other transitions. In these cases, {\sc Minfit} allows for one to mask pixels or pixel regions that then contribute no information to the fit.

\begin{figure*}[htb]
\figurenum{6} \centering
\includegraphics[width=0.95\textwidth]{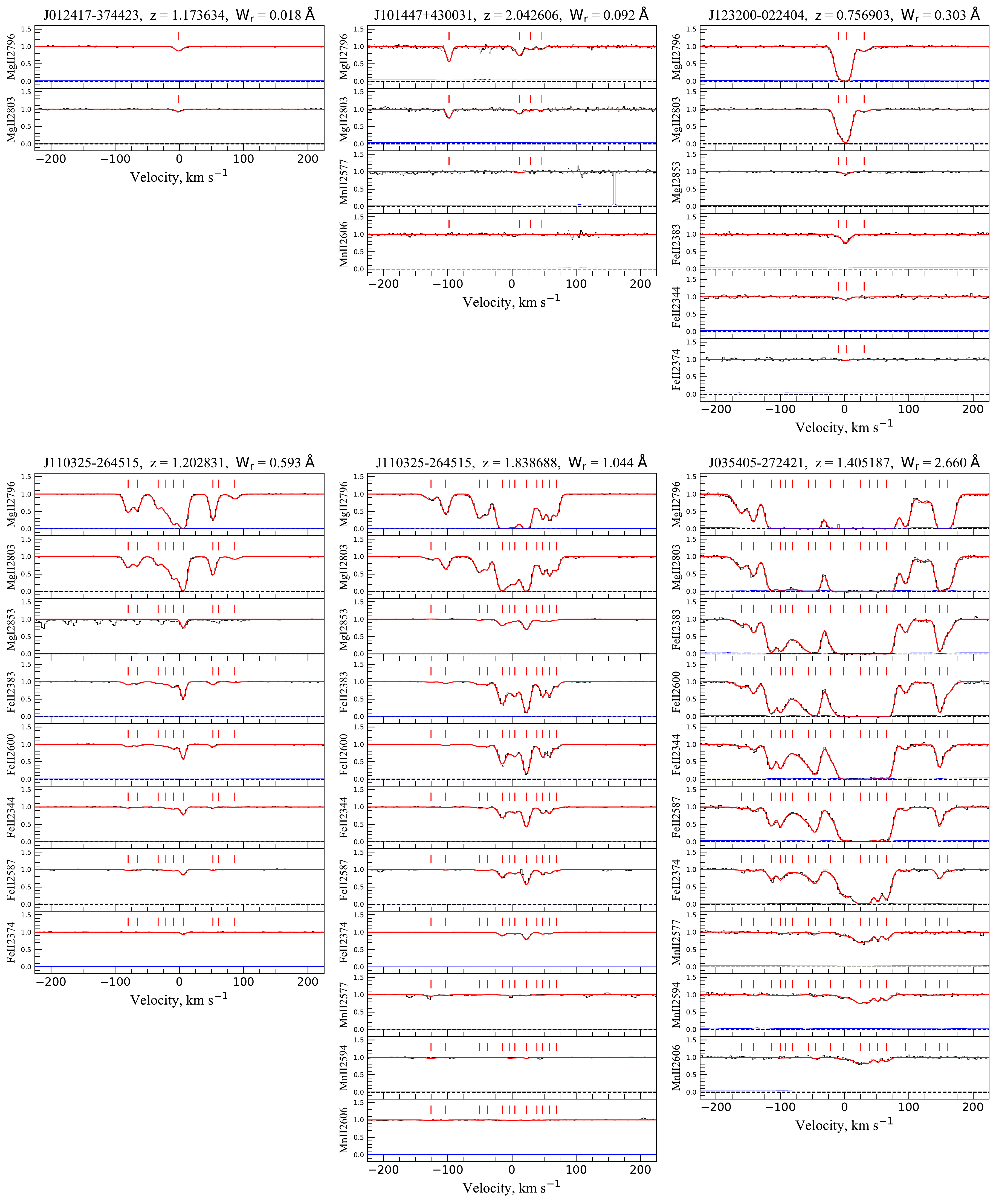}
\caption{Six selected systems illustrating the range of absorption profiles strengths and kinematic complexity. The VP models (red) are superimposed over the data (blue). Ticks above the spectra indicate the VP component velocity centroids, which appear above each transition whether or not absorption was detected at the $3\sigma$ significance level. If a given associated transition (see Table~\ref{tab:transitions}) is not presented with a given system, it is because it was not covered by the spectrum. The VP fitted parameters for these systems are listed in Table~\ref{tab:sampleminfitnumbers}. The complete figure set (422 images) is available in the online journal.}
\label{fig:profile1}
\end{figure*}

Finding physically meaningful VP fits in regions of extreme line saturation can be very challenging.  For example, consider the velocity region $\simeq 0\,$--75~{\kms} in the $z_{\rm abs}=1.405187$ system in the quasar spectrum of J035405--272421 illustrated in Figure~\ref{fig:profile1}, where both the {\MgIIdblt} and the majority of the {\FeII} transitions are highly saturated. In extreme cases such as this, the least-squares fitting engine can fixate on a local $\chi^2$ minimum in which the ratio of $N({\FeII})/N({\MgII})$ is unphysical, as it is astrophysically rare for [Mg/Fe] to fall an order of magnitude {\it below\/} the solar value. In their VP decomposition of some two-dozen {\MgII} absorption systems observed with the HIRES spectrograph, \citet{churchill03} found the relation 
\begin{equation}
    \log N({\FeII}) = 1.37 \, \log\! N({\MgII}) \, - \, 0.41 \, ,
\end{equation} 
for the unsaturated velocity regions of the absorption profiles. For our work here, we constrain the {\FeII} column densities to obey this relation in highly saturated velocity regions.  The user specifies these velocity ranges when the constraint is deemed necessary. We find that this successfully prevents unphysical {\FeII} to {\MgII} column density ratios in these specified velocity regions.

As illustrated in Figure~\ref{fig:flowchart}, once the initial model is constructed, any bad pixels are masked, and any saturated velocity regions are specified, {\sc Minfit} performs a refinement of the model by minimizing the $\chi^2$ statistic. Essentially, {\sc Minfit} is a driver that sets up $M$ nonlinear functions in $n$ parameters and feeds the vector of functions to the netlib.org routine {\sc dnls1}  \citep{more78}. The $M$ functions are the individual terms of the $\chi^2$ statistic, one for each pixel for all transitions, and the $n$ parameters are the VP free parameters, where the value of $n$ depends on the number of components, transitions, and ions. The least squares fit (the box labeled ``LSF" in Figure~\ref{fig:flowchart}) is performed by the routine {\sc dnls1}, a modification of the Levenberg-Marquardt algorithm.  Two of its main characteristics involve the proper use of implicitly scaled parameters and an optimal choice for the correction terms. The routine approximates the Jacobian by forward differencing. The use of implicitly scaled parameters achieves scale invariance and limits the size of the correction in any direction where the functions are changing rapidly. The optimal choice of the correction guarantees (under reasonable conditions) global convergence from starting points far from the initial guess solution and a fast rate of convergence for problems with small residuals. Further details, including the methods for computing the uncertainties in the fitted parameters are described in  \citet{churchill97}.  

Following the LSF, {\sc Minfit} computes the errors in the VP parameters, $\sigma_N$, $\sigma _b$, and $\sigma _z$ (recall that each component is fitted in redshift space and later converted to rest-frame velocity). The robustness of components is examined in two ways. First, for each ion, each component, $i$, is checked against it nearest neighbor components, $i\!-\!1$ and $i\!+\!1$, for redshift overlap using the conditions $\left| (z_i\!-\! z_{i-1})/\sigma _{z_i} \right| \! > \! 1$ or 
$\left| (z_i \!-\! z_{i+1})/\sigma _{z_i}\right| \! >\! 1 $.
If a component satisfies one of those conditions, it is flagged for significance testing, as described below. For the second robustness check, a ``badness" parameter is computed for each component,
\begin{equation}
\mathrm{badness} = \left[
\bigg({\sigma_N \over N}\bigg)^2 +
\bigg({\sigma_b \over b}\bigg)^2 + 
\bigg({\sigma_z \over z}\bigg)^2 \,
\right]^{1/2} \, .
\end{equation}


\startlongtable

\begin{deluxetable*}{rlrlrlrlrlr}
\tabletypesize{\scriptsize}
\tablewidth{0pt}
\tablecolumns{11}
\tablecaption{Voigt Profile Fitted Parameters\label{tab:sampleminfitnumbers}}
\tablehead{
\colhead{\phantom{x}} & \multicolumn{2}{c}{\MgII} &
\multicolumn{2}{c}{\FeII} & \multicolumn{2}{c}{\MgI} &
\multicolumn{2}{c}{\MnII} & \multicolumn{2}{c}{\CaII}\\[-0.2cm]
\colhead{$v$} & \colhead{$\log N$} & \colhead{$b$} & 
\colhead{$\log N$} & \colhead{$b$} & \colhead{$\log N$} & \colhead{$b$} & 
\colhead{$\log N$} & \colhead{$b$} & \colhead{$\log N$} & \colhead{$b$}\\[-0.3cm]
\colhead{[km s$^{-1}$]} & \colhead{[cm$^{-2}$]} & \colhead{[km s$^{-1}$]} & 
\colhead{[cm$^{-2}$]} & \colhead{[km s$^{-1}$]} & \colhead{[cm$^{-2}$]} & \colhead{[km s$^{-1}$]} & 
\colhead{[cm$^{-2}$]} & \colhead{[km s$^{-1}$]} & \colhead{[cm$^{-2}$]} & \colhead{[km s$^{-1}$]}
}
\startdata
\\[-5pt]
\multicolumn{11}{c}{J012417$-$374423, $z_{\rm abs}=1.173635$, $N_{\hbox{\tiny VP}}=1$, ${\rm badness}= 1.5$, ${\rm CL}=97$\%}\\
$-0.91$ & $11.64$$\pm$0.03 & $6.92$$\pm$0.67 & $\leq10.62$ & \multicolumn{1}{c}{\nodata} & $\leq10.55$ & \multicolumn{1}{c}{\nodata} & $\leq10.61$ & \multicolumn{1}{c}{\nodata}& \multicolumn{1}{c}{\nodata}& \multicolumn{1}{c}{\nodata}\\
\multicolumn{11}{c}{J101447$+$430031, $z_{\rm abs}=2.042606$, $N_{\hbox{\tiny VP}}=4$, ${\rm badness}= 1.5$, ${\rm CL}=97$\%}\\
$-98.07$ & $12.11$$\pm$0.02 & $3.73$$\pm$0.33 & $\leq10.75$ & \multicolumn{1}{c}{\nodata} &\multicolumn{1}{c}{\nodata} &\multicolumn{1}{c}{\nodata} & $\leq10.83$ & \multicolumn{1}{c}{\nodata}& \multicolumn{1}{c}{\nodata}& \multicolumn{1}{c}{\nodata}\\[-3pt]
$11.34$ & $11.91$$\pm$0.03 & $5.76$$\pm$0.67 & $\leq10.84$ & \multicolumn{1}{c}{\nodata} &\multicolumn{1}{c}{\nodata} &\multicolumn{1}{c}{\nodata} & $11.29$$\pm$0.25 & $5.76$$\pm$0.67& \multicolumn{1}{c}{\nodata}& \multicolumn{1}{c}{\nodata}\\[-3pt]
$29.01$ & $11.35$$\pm$0.11 & $4.52$$\pm$2.29 & $\leq10.78$ & \multicolumn{1}{c}{\nodata} &\multicolumn{1}{c}{\nodata} &\multicolumn{1}{c}{\nodata} & $\leq10.87$ & \multicolumn{1}{c}{\nodata}& \multicolumn{1}{c}{\nodata}& \multicolumn{1}{c}{\nodata}\\[-3pt]
$45.26$ & $11.33$$\pm$0.11 & $4.87$$\pm$2.39 & $\leq10.78$ & \multicolumn{1}{c}{\nodata} &\multicolumn{1}{c}{\nodata} &\multicolumn{1}{c}{\nodata} & $\leq10.96$ & \multicolumn{1}{c}{\nodata}& \multicolumn{1}{c}{\nodata}& \multicolumn{1}{c}{\nodata}\\
\multicolumn{11}{c}{J123200$-$022404, $z_{\rm abs}=0.756903$, $N_{\hbox{\tiny 
VP}}=3$, ${\rm badness}= 1.5$, ${\rm CL}=97$\%}\\
$-9.28$ & $12.75$$\pm$0.07 & $6.24$$\pm$0.51 & $11.58$$\pm$0.17 & $6.24$$\pm$0.51 & $\leq10.72$ & \multicolumn{1}{c}{\nodata}& \multicolumn{1}{c}{\nodata}& \multicolumn{1}{c}{\nodata}& \multicolumn{1}{c}{\nodata}& \multicolumn{1}{c}{\nodata}\\[-3pt]
$2.42$ & $13.33$$\pm$0.02 & $6.73$$\pm$0.23 & $12.30$$\pm$0.03 & $6.73$$\pm$0.23 & $10.87$$\pm$0.08 & $6.73$$\pm$0.23& \multicolumn{1}{c}{\nodata}& \multicolumn{1}{c}{\nodata}& \multicolumn{1}{c}{\nodata}& \multicolumn{1}{c}{\nodata}\\[-3pt]
$30.47$ & $11.77$$\pm$0.04 & $9.08$$\pm$1.18 & $11.23$$\pm$0.34 & $9.08$$\pm$1.18 & $\leq10.78$ & \multicolumn{1}{c}{\nodata}& \multicolumn{1}{c}{\nodata}& \multicolumn{1}{c}{\nodata}& \multicolumn{1}{c}{\nodata}& \multicolumn{1}{c}{\nodata}\\
\multicolumn{11}{c}{J110325$-$264515, $z_{\rm abs}=1.202831$, $N_{\hbox{\tiny VP}}=9$, ${\rm badness}= 1.5$, ${\rm CL}=97$\%}\\
$-79.98$ & $12.34$$\pm$0.01 & $5.84$$\pm$0.01 & $11.80$$\pm$0.02 & $5.84$$\pm$0.01 & $\leq11.56$ & \multicolumn{1}{c}{\nodata} & $\leq10.47$ & \multicolumn{1}{c}{\nodata} & $\leq10.15$ & \multicolumn{1}{c}{\nodata}\\[-3pt]
$-65.77$ & $12.21$$\pm$0.01 & $5.12$$\pm$0.01 & $11.58$$\pm$0.04 & $5.12$$\pm$0.01 & $\leq11.63$ & \multicolumn{1}{c}{\nodata} & $\leq10.47$ & \multicolumn{1}{c}{\nodata} & $\leq10.10$ & \multicolumn{1}{c}{\nodata}\\[-3pt]
$-33.23$ & $12.19$$\pm$0.01 & $5.85$$\pm$0.01 & $11.49$$\pm$0.03 & $5.85$$\pm$0.01 & $\leq11.67$ & \multicolumn{1}{c}{\nodata} & $\leq10.47$ & \multicolumn{1}{c}{\nodata} & $\leq10.07$ & \multicolumn{1}{c}{\nodata}\\[-3pt]
$-22.33$ & $12.02$$\pm$0.08 & $3.08$$\pm$0.41 & $11.50$$\pm$0.06 & $3.08$$\pm$0.41 & $\leq11.61$ & \multicolumn{1}{c}{\nodata} & $\leq10.37$ & \multicolumn{1}{c}{\nodata} & $\leq9.97$ & \multicolumn{1}{c}{\nodata}\\[-3pt]
$-9.16$ & $12.84$$\pm$0.01 & $7.33$$\pm$0.01 & $12.14$$\pm$0.01 & $7.33$$\pm$0.01 & $\leq10.49$ & \multicolumn{1}{c}{\nodata} & $\leq10.54$ & \multicolumn{1}{c}{\nodata} & $\leq10.15$ & \multicolumn{1}{c}{\nodata}\\[-3pt]
$5.80$ & $15.06$$\pm$0.02 & $2.67$$\pm$0.01 & $12.55$$\pm$0.01 & $2.67$$\pm$0.01 & $\simeq 11.3$ & $2.67$$\pm$0.01 & $\leq10.37$ & \multicolumn{1}{c}{\nodata} & $\leq10.00$ & \multicolumn{1}{c}{\nodata}\\[-3pt]
$52.08$ & $12.57$$\pm$0.01 & $4.02$$\pm$0.01 & $11.70$$\pm$0.03 & $4.02$$\pm$0.01 & $\leq11.62$ & \multicolumn{1}{c}{\nodata} & $\leq10.43$ & \multicolumn{1}{c}{\nodata} & $\leq10.07$ & \multicolumn{1}{c}{\nodata}\\[-3pt]
$61.25$ & $11.13$$\pm$0.07 & $6.72$$\pm$0.78 & $\leq10.50$ & \multicolumn{1}{c}{\nodata} & $\leq11.73$ & \multicolumn{1}{c}{\nodata} & $\leq10.51$ & \multicolumn{1}{c}{\nodata} & $\leq10.15$ & \multicolumn{1}{c}{\nodata}\\[-3pt]
$86.15$ & $11.70$$\pm$0.02 & $7.36$$\pm$0.45 & $11.07$$\pm$0.12 & $7.36$$\pm$0.45 & $\leq11.58$ & \multicolumn{1}{c}{\nodata} & $\leq10.54$ & \multicolumn{1}{c}{\nodata} & $\leq10.17$ & \multicolumn{1}{c}{\nodata}\\
\multicolumn{11}{c}{J110325$-$264515, $z_{\rm abs}=1.838689$, $N_{\hbox{\tiny VP}}=12$, ${\rm badness}= 1.5$, ${\rm CL}=99$\%}\\
$-126.13$ & $11.86$$\pm$0.01 & $9.10$$\pm$0.28 & $10.92$$\pm$0.10 & $9.10$$\pm$0.28 & $\leq10.44$ & \multicolumn{1}{c}{\nodata} & $11.48$$\pm$0.03 & $9.10$$\pm$0.28& \multicolumn{1}{c}{\nodata}& \multicolumn{1}{c}{\nodata}\\[-3pt]
$-103.14$ & $12.41$$\pm$0.01 & $6.12$$\pm$0.06 & $11.54$$\pm$0.02 & $6.12$$\pm$0.06 & $\leq10.38$ & \multicolumn{1}{c}{\nodata} & $10.99$$\pm$0.06 & $6.12$$\pm$0.06& \multicolumn{1}{c}{\nodata}& \multicolumn{1}{c}{\nodata}\\[-3pt]
$-50.66$ & $12.61$$\pm$0.01 & $7.69$$\pm$0.09 & $11.87$$\pm$0.01 & $7.69$$\pm$0.09 & $10.54$$\pm$0.06 & $7.69$$\pm$0.09 & $10.56$$\pm$0.18 & $7.69$$\pm$0.09& \multicolumn{1}{c}{\nodata}& \multicolumn{1}{c}{\nodata}\\[-3pt]
$-38.04$ & $12.26$$\pm$0.01 & $4.04$$\pm$0.11 & $11.64$$\pm$0.02 & $4.04$$\pm$0.11 & $\leq10.45$ & \multicolumn{1}{c}{\nodata} & $\leq10.29$ & \multicolumn{1}{c}{\nodata}& \multicolumn{1}{c}{\nodata}& \multicolumn{1}{c}{\nodata}\\[-3pt]
$-15.10$ & $13.50$$\pm$0.01 & $6.62$$\pm$0.03 & $12.89$$\pm$0.01 & $6.62$$\pm$0.03 & $11.33$$\pm$0.01 & $6.62$$\pm$0.03 & $10.75$$\pm$0.11 & $6.62$$\pm$0.03& \multicolumn{1}{c}{\nodata}& \multicolumn{1}{c}{\nodata}\\[-3pt]
$-3.50$ & $12.69$$\pm$0.05 & $1.61$$\pm$0.07 & $11.80$$\pm$0.02 & $1.61$$\pm$0.07 & $10.61$$\pm$0.05 & $1.61$$\pm$0.07 & $11.01$$\pm$0.05 & $1.61$$\pm$0.07& \multicolumn{1}{c}{\nodata}& \multicolumn{1}{c}{\nodata}\\[-3pt]
$4.58$ & $12.95$$\pm$0.01 & $5.66$$\pm$0.07 & $12.55$$\pm$0.01 & $5.66$$\pm$0.07 & $10.75$$\pm$0.04 & $5.66$$\pm$0.07 & $11.07$$\pm$0.05 & $5.66$$\pm$0.07& \multicolumn{1}{c}{\nodata}& \multicolumn{1}{c}{\nodata}\\[-3pt]
$22.46$ & $13.82$$\pm$0.01 & $5.21$$\pm$0.02 & $13.19$$\pm$0.01 & $5.21$$\pm$0.02 & $11.50$$\pm$0.01 & $5.21$$\pm$0.02 & $11.23$$\pm$0.03 & $5.21$$\pm$0.02& \multicolumn{1}{c}{\nodata}& \multicolumn{1}{c}{\nodata}\\[-3pt]
$38.69$ & $11.95$$\pm$0.01 & $1.50$$\pm$0.09 & $11.42$$\pm$0.03 & $1.50$$\pm$0.09 & $10.23$$\pm$0.09 & $1.50$$\pm$0.09 & $10.53$$\pm$0.14 & $1.50$$\pm$0.09& \multicolumn{1}{c}{\nodata}& \multicolumn{1}{c}{\nodata}\\[-3pt]
$47.82$ & $12.53$$\pm$0.01 & $3.16$$\pm$0.07 & $12.48$$\pm$0.01 & $3.16$$\pm$0.07 & $10.74$$\pm$0.03 & $3.16$$\pm$0.07 & $10.57$$\pm$0.14 & $3.16$$\pm$0.07& \multicolumn{1}{c}{\nodata}& \multicolumn{1}{c}{\nodata}\\[-3pt]
$58.56$ & $12.57$$\pm$0.01 & $2.87$$\pm$0.06 & $12.41$$\pm$0.01 & $2.87$$\pm$0.06 & $10.65$$\pm$0.04 & $2.87$$\pm$0.06 & $\leq10.25$ & \multicolumn{1}{c}{\nodata}& \multicolumn{1}{c}{\nodata}& \multicolumn{1}{c}{\nodata}\\[-3pt]
$69.13$ & $12.41$$\pm$0.01 & $4.41$$\pm$0.09 & $11.92$$\pm$0.01 & $4.41$$\pm$0.09 & $10.30$$\pm$0.09 & $4.41$$\pm$0.09 & $\leq10.30$ & \multicolumn{1}{c}{\nodata}& \multicolumn{1}{c}{\nodata}& \multicolumn{1}{c}{\nodata}\\
\multicolumn{11}{c}{J035405$-$272421, $z_{\rm abs}=1.405188$, $N_{\hbox{\tiny VP}}=18$, ${\rm badness}= 1.5$, ${\rm CL}=49$\% }\\
$-160.39$ & $12.55$$\pm$0.02 & $10.91$$\pm$0.56 & $12.33$$\pm$0.03 & $10.91$$\pm$0.56 & \multicolumn{1}{c}{\nodata} & \multicolumn{1}{c}{\nodata} & $\leq11.04$ & \multicolumn{1}{c}{\nodata}& \multicolumn{1}{c}{\nodata}& \multicolumn{1}{c}{\nodata}\\[-3pt]
$-141.33$ & $12.64$$\pm$0.02 & $5.63$$\pm$0.28  & $12.50$$\pm$0.02 & $5.63$$\pm$0.28  & \multicolumn{1}{c}{\nodata} & \multicolumn{1}{c}{\nodata} & $10.99$$\pm$0.32 & $5.63$$\pm$0.28& \multicolumn{1}{c}{\nodata}& \multicolumn{1}{c}{\nodata}\\[-3pt]
$-113.89$ & $13.35$$\pm$0.04 & $6.36$$\pm$0.16  & $13.37$$\pm$0.01 & $6.36$$\pm$0.16  & \multicolumn{1}{c}{\nodata} & \multicolumn{1}{c}{\nodata} & $10.83$$\pm$0.49 & $6.36$$\pm$0.16& \multicolumn{1}{c}{\nodata}& \multicolumn{1}{c}{\nodata}\\[-3pt]
$-99.75$  & $13.38$$\pm$0.23 & $2.78$$\pm$0.22  & $13.21$$\pm$0.02 & $2.78$$\pm$0.22  & \multicolumn{1}{c}{\nodata} & \multicolumn{1}{c}{\nodata} & $10.98$$\pm$0.30 & $2.78$$\pm$0.22& \multicolumn{1}{c}{\nodata}& \multicolumn{1}{c}{\nodata}\\[-3pt]
$-91.97$  & $12.52$$\pm$0.26 & $2.33$$\pm$1.01  & $12.51$$\pm$0.08 & $2.33$$\pm$1.01  & \multicolumn{1}{c}{\nodata} & \multicolumn{1}{c}{\nodata} & $10.52$$\pm$0.90 & $2.33$$\pm$1.01& \multicolumn{1}{c}{\nodata}& \multicolumn{1}{c}{\nodata}\\[-3pt]
$-80.97$  & $13.74$$\pm$0.08 & $25.95$$\pm$3.39 & $13.32$$\pm$0.08 & $25.95$$\pm$3.39 & \multicolumn{1}{c}{\nodata} & \multicolumn{1}{c}{\nodata} & $11.64$$\pm$0.19 & $25.95$$\pm$3.39& \multicolumn{1}{c}{\nodata}& \multicolumn{1}{c}{\nodata}\\[-3pt]
$-56.41$  & $13.71$$\pm$0.23 & $11.32$$\pm$3.22 & $13.46$$\pm$0.23 & $11.32$$\pm$3.22 & \multicolumn{1}{c}{\nodata} & \multicolumn{1}{c}{\nodata} & $\leq10.95$ & \multicolumn{1}{c}{\nodata}& \multicolumn{1}{c}{\nodata}& \multicolumn{1}{c}{\nodata}\\[-3pt]
$-45.25$  & $13.59$$\pm$0.21 & $6.76$$\pm$0.45  & $13.54$$\pm$0.09 & $6.76$$\pm$0.45  & \multicolumn{1}{c}{\nodata} & \multicolumn{1}{c}{\nodata} & $11.53$$\pm$0.12 & $6.76$$\pm$0.45& \multicolumn{1}{c}{\nodata}& \multicolumn{1}{c}{\nodata}\\[-3pt]
$-21.63$  & $13.19$$\pm$0.07 & $5.87$$\pm$0.53  & $12.89$$\pm$0.07 & $5.87$$\pm$0.53  & \multicolumn{1}{c}{\nodata} & \multicolumn{1}{c}{\nodata} & $10.67$$\pm$0.64 & $5.87$$\pm$0.53& \multicolumn{1}{c}{\nodata}& \multicolumn{1}{c}{\nodata}\\[-3pt]
$-1.43$   & $14.53$$\pm$0.05 & $10.13$$\pm$0.72 & $14.30$$\pm$0.05 & $10.13$$\pm$0.72 & \multicolumn{1}{c}{\nodata} & \multicolumn{1}{c}{\nodata} & $11.92$$\pm$0.07 & $10.13$$\pm$0.72& \multicolumn{1}{c}{\nodata}& \multicolumn{1}{c}{\nodata}\\[-3pt]
$24.38$   & $15.09$$\pm$0.05 & $13.06$$\pm$1.61 & $14.86$$\pm$0.05 & $13.06$$\pm$1.61 & \multicolumn{1}{c}{\nodata} & \multicolumn{1}{c}{\nodata} & $12.57$$\pm$0.05 & $13.06$$\pm$1.61& \multicolumn{1}{c}{\nodata}& \multicolumn{1}{c}{\nodata}\\[-3pt]
$38.83$   & $14.65$$\pm$0.11 & $4.56$$\pm$0.93  & $14.42$$\pm$0.11 & $4.56$$\pm$0.93  & \multicolumn{1}{c}{\nodata} & \multicolumn{1}{c}{\nodata} & $11.94$$\pm$0.16 & $4.56$$\pm$0.93& \multicolumn{1}{c}{\nodata}& \multicolumn{1}{c}{\nodata}\\[-3pt]
$51.57$   & $15.60$$\pm$0.39 & $1.83$$\pm$0.24  & $15.38$$\pm$0.39 & $1.83$$\pm$0.24  & \multicolumn{1}{c}{\nodata} & \multicolumn{1}{c}{\nodata} & $11.97$$\pm$0.04 & $1.83$$\pm$0.24& \multicolumn{1}{c}{\nodata}& \multicolumn{1}{c}{\nodata}\\[-3pt]
$65.15$   & $15.10$$\pm$0.05 & $7.02$$\pm$0.09  & $14.21$$\pm$0.01 & $7.02$$\pm$0.09  & \multicolumn{1}{c}{\nodata} & \multicolumn{1}{c}{\nodata} & $12.13$$\pm$0.03 & $7.02$$\pm$0.09& \multicolumn{1}{c}{\nodata}& \multicolumn{1}{c}{\nodata}\\[-3pt]
$94.62$ & $12.98$$\pm$0.01 & $6.47$$\pm$0.16 & $12.53$$\pm$0.01 & $6.47$$\pm$0.16 & \multicolumn{1}{c}{\nodata} & \multicolumn{1}{c}{\nodata} & $\leq10.95$ & \multicolumn{1}{c}{\nodata}& \multicolumn{1}{c}{\nodata}& \multicolumn{1}{c}{\nodata}\\[-3pt]
$125.71$ & $12.23$$\pm$0.04 & $14.33$$\pm$1.81 & $11.79$$\pm$0.09 & $14.33$$\pm$1.81 & \multicolumn{1}{c}{\nodata} & \multicolumn{1}{c}{\nodata} & $11.24$$\pm$0.25 & $14.33$$\pm$1.81& \multicolumn{1}{c}{\nodata}& \multicolumn{1}{c}{\nodata}\\[-3pt]
$148.04$ & $13.64$$\pm$0.06 & $4.44$$\pm$0.13 & $13.25$$\pm$0.01 & $4.44$$\pm$0.13 & \multicolumn{1}{c}{\nodata} & \multicolumn{1}{c}{\nodata} & $\leq10.96$ & \multicolumn{1}{c}{\nodata}& \multicolumn{1}{c}{\nodata}& \multicolumn{1}{c}{\nodata}\\[-3pt]
$159.68$ & $13.33$$\pm$0.04 & $8.97$$\pm$0.43 & $12.68$$\pm$0.03 & $8.97$$\pm$0.43 & \multicolumn{1}{c}{\nodata} & \multicolumn{1}{c}{\nodata} & $\leq10.95$ & \multicolumn{1}{c}{\nodata}& \multicolumn{1}{c}{\nodata}& \multicolumn{1}{c}{\nodata}\\[2pt]
\enddata
\tablecomments{Table~\ref{tab:sampleminfitnumbers} is published in its entirety in machine-readable format. A portion is shown here for guidance regarding its form and content.}
\end{deluxetable*}

If any components have a badness parameter that exceed a user specified value, and/or any components are identified to overlap in redshift, then a series of significance tests are instigated. We have adopted a default value of ${\rm badness} = 1.5$, but have relaxed this parameter as needed for various systems (typically those that exhibit extreme saturation). If no VP components exceed the badness threshold and none overlap their nearest neighbor, the current VP model is adopted.  

Components flagged for significance checking are rank ordered with priority given to redshift overlap followed by a ranking from highest badness to lowest badness.  The highest ranked flagged component is simply deleted from the ``current" VP model and an LSF is obtained for this new ``test" model, which has one fewer VP component.  Accounting for the different degrees of freedom in the two models, an $F$-test is performed between the ``current" and ``test" models to determine whether inclusion of the flagged component provides a statistically significant improvement in the $\chi^2$ statistic to a user specified confidence level.  We adopt a default confidence level of 97\%, though we have relaxed this number for selected systems as needed. No systems have been fitted for which the confidence level of the components is below 90\%.

If the tested component was statistically significant, {\sc Minfit} retains the ``current" VP model and then proceeds to investigate the statistical significance of the next line in the sorted ``bad" component array, and so on until the array is exhausted.  If the tested component was not statistically significant, then the ``test" model is adopted as the ``current" model; the number of components required to model the absorption system is now reduced. {\sc Minfit} then identifies whether any neighboring components exhibit redshift overlap and computes the badness parameters for this newly adopted ``current" VP model and proceeds to test the significance of the components.  The entire process is automated and objective.  A final VP model is adopted when all components are determined to be statistically significant. Human manipulation of the objective outcome of the process (i.e., the adopted VP model) can occur only via modification of the threshold badness parameter and/or the threshold confidence level. Those systems for which we modified the default values are noted in the descriptions of individual systems in Appendix~\ref{sec:appendix}.

Once the final VP model is adopted and the final uncertainties have been computed, upper limits are computed for the component column densities for associated  ions  (i.e., {\MgI}, {\FeII}, {\MnII}, {\CaII}) for which no transitions were formally detected at the velocity positions of {\MgII} components. The Doppler $b$ parameter of the specific {\MgII} component is adopted and the column density limit is determined from the $3\sigma$ equivalent width detection limit across the velocity range of the component

We note that {\sc Minfit} is a deterministic algorithm, in that, for a given input model and user specified parameters (masking, saturated regions, badness, and confidence level), the computational path of the least-squares fit and the final solution will always be the same.  By exploring the outcomes as a function of variations in the input model and/or the user specified parameters, we inspected various final models. In the end, the adopted final model for a given system is a human decision \citep[however, see][]{bainbridge17a}. 

\section{Results and Discussion}
\label{sec:results}

We have obtained the VP models of 422 {\MgII}-selected absorption systems in our kinematic sample.  These models provide the number of components $N_{\hbox{\tiny VP}}$, their column densities, $N$, Doppler $b$ parameters, and rest-frame velocities.  Our modeling yielded a total of 2989 components. We thus have estimates of the number of ``clouds", the product of their average ionic number densities and the line-of-sight depth of the ``cloud", an estimate of their kinematic and/or thermal broadening, and their line-of-sight projected relative rest-frame kinematics, assuming the ``clouds" are spatially distinct entities.


In Figure~\ref{fig:profile1}, we present six selected absorption systems. These six representative systems are presented to display the dynamic range in their properties, from simple single-component weak absorbers for which only the {\MgII} doublet is covered (J012417$-$374423, $z_{\rm abs}\!=\!1.173634$) to highly complex multi-component absorbers for which several associated transitions are detected and/or covered (J035405$-$272421, $z_{\rm abs}\!=\!1.405187$). The red curves through the data (blue) are Voigt profile (VP) models of the absorption (see~Section~\ref{sec:vpanalysis}) and the vertical ticks above the continuum provide the velocities of the individual VP components. The VP fitted parameters for these systems are listed in Table~\ref{tab:sampleminfitnumbers}. The VP modeling will be described in detail in Section~\ref{sec:vpanalysis}. The complete figure set (422 images) is available in the online journal.

To illustrate these data and their typical uncertainties, we present the VP fitted parameters in Table~\ref{tab:sampleminfitnumbers} for the six selected absorption systems shown in Figure~\ref{fig:profile1}. Column (1) tabulates the rest-frame velocity of the component. Columns (2)--(11) tabulate the column densities, or their upper limits, and the Doppler $b$ parameters for each component for the {\MgII}, {\FeII}, {\MgI}, {\MnII}, and {\CaII} ions, respectively. Systems that are present in our sample that were also VP modelled by \citet{churchill97} and \citet{churchill03} have been refitted with the methods described in Section~\ref{sec:vpanalysis} for uniformity. Table~\ref{tab:sampleminfitnumbers} is published in its entirety in machine-readable format. A portion is shown here for guidance regarding its form and content.

\subsection{VP Components Line Density}
\label{subsec:nvp}

\begin{figure*}[tb]
\figurenum{7} \centering
\includegraphics[width=0.8\textwidth]{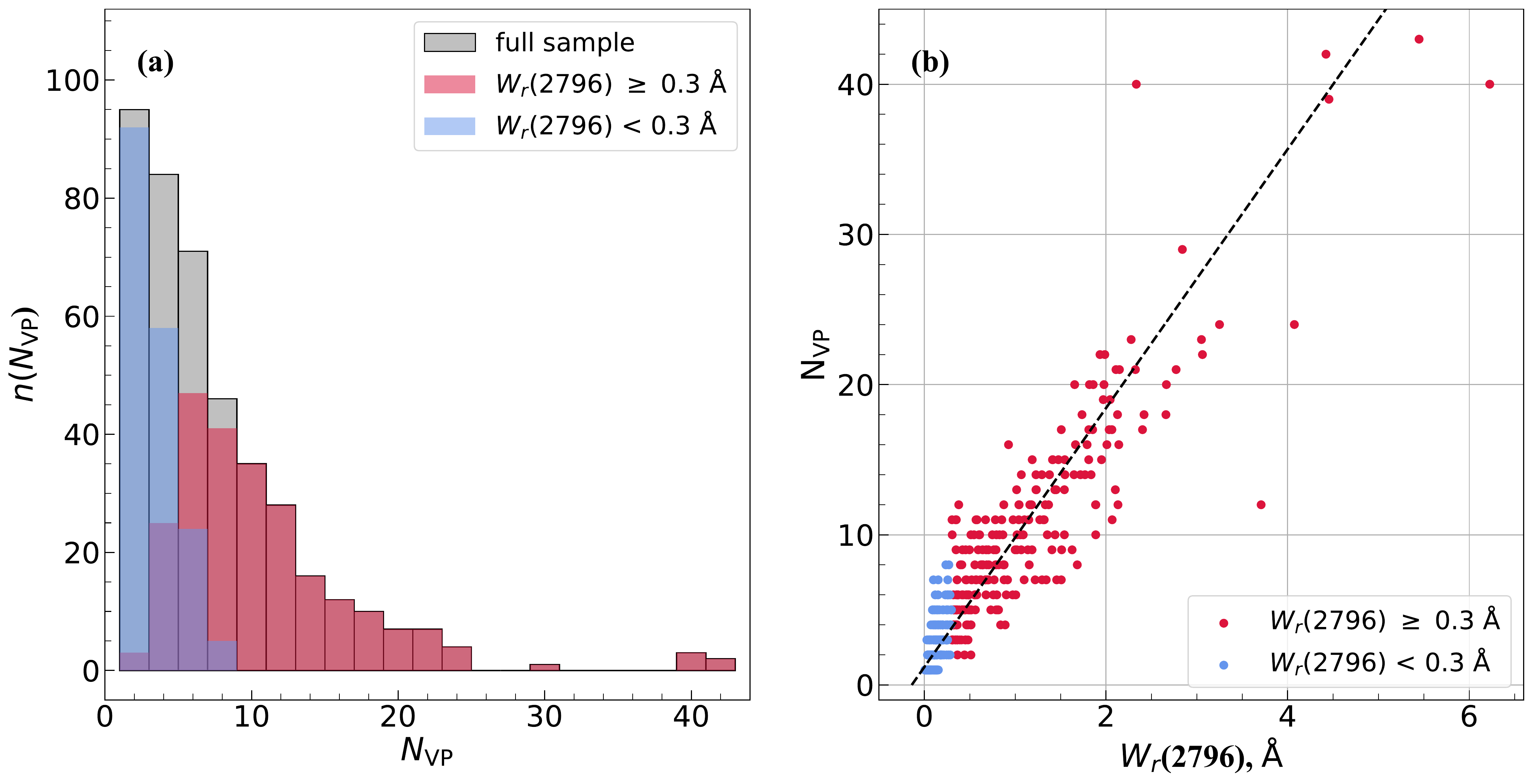}
\caption{(a) The distributions of the number of VP components $N_{\hbox{\tiny VP}}$ for $W_r(2796) < 0.3$ {\AA} (blue), $W_r(2796) \geq 0.3$ {\AA} (pink), and the full sample (gray); overlap of the weak and strong systems appears as purple. (b) The number of components, $N_{\hbox{\tiny VP}}$, versus the {\MgII}~$\lambda 2796$ rest-frame equivalent width for weak (blue points) and strong (pink) systems.  The dotted line indicates a linear least-squares fit having a slope of 8.62 clouds~{\AA}$^{-1}$.} 
\label{fig:nclouds}
\end{figure*}

In Figure~\ref{fig:nclouds}(a) we show the binned distributions of the number of VP components for all systems, weak systems ($W_r(2796) < 0.3$ {\AA}), and strong systems ($W_r(2796) \geq 0.3$ {\AA}); the color scheme of the histograms is the same as used in Figure~\ref{fig:zabs}. The full sample was modeled with an average of $\langle N_{\hbox{\tiny VP}} \rangle = 7.1$ components, whereas the weak systems were modeled with an average of $\langle N_{\hbox{\tiny VP}} \rangle = 2.7$ components and the strong systems with $\langle N_{\hbox{\tiny VP}} \rangle = 10.3$ components. 

\citet{churchill97} modeled simulated multi-component {\MgII} absorption systems in synthetic spectra having the characteristics of HIRES/Keck spectra and found that on average $\sim\!\!30$\% of VP components were not recovered using {\sc Minfit}. Those simulated profiles were generated using the observed distributions of VP component velocity separations, column densities, and Doppler $b$ parameters from the VP decomposition of two-dozen systems observed with Keck/HIRES, but the number of components was fixed at $N_{\hbox{\tiny VP}} = 10$ as a control condition. The signal-to-noise ratio for a given simulation was also held fixed; the quoted results here are for a simulation such that the $5\sigma$ equivalent width detection threshold was 0.02~{\AA}. The average number of components recovered in these experiments did vary with signal-to-noise ratio and the presence (or non-presence) of associated transitions with clear kinematic structure. Component recovery improved to $\sim\!90$\% with the presence of associated transitions and always decreased as signal-to-noise ratio decreased.  Though the tests are based on the assumption that {\MgII} absorption profiles are a complex of VP components (a clearly simplistic scenario), if the outcomes can be applied directly to our sample, it would suggest that the actual mean numbers of components are $\sim\!\!11$--43\% higher than what we report here.

In Figure \ref{fig:nclouds}(b), the number of VP components is plotted as a function of the {\MgII}~$\lambda 2796$ rest-frame system equivalent width.  Blue points represent weak systems and pink points represent strong systems.  A linear fit to the full sample resulted in a slope of $8.62\pm0.23$~clouds~{\AA}$^{-1}$, which can be interpreted as the VP component line density. Note the increased scatter for $W_r(2796) > 2$~{\AA}, where VP fitting can become challenging and problematic for highly saturated or partially saturated absorption profiles. For $W_r(2796) \leq 2$~{\AA}, the absolute standard deviation of $N_{\hbox{\tiny VP}}$ about the fitted relation is $3.2$ ``clouds".

Previous work examined the inverse of the VP component density, i.e., the slope in terms of {\AA}~cloud$^{-1}$.  For our sample, our fit corresponds to $0.116\pm0.003$~{\AA}~cloud$^{-1}$.  The slope found by \citet{churchill97} for a sample of 36 {\MgII} systems observed with Keck/HIRES (the same spectral resolution as this study) was $0.076\pm0.004$~{\AA}~cloud$^{-1}$.  Their higher number of components per unit equivalent width (13.2~clouds~{\AA}$^{-1}$) is likely due to the modifications we made to the {\sc Minfit} program. The distribution of the signal-to-noise ratio of the spectra clearly play a role, as higher quality data can constrain the VP models to have a larger number of ``clouds"; however, the distributions of our survey and that of \citet{churchill97} are statistically consistent. In \citet{churchill97}, the ``badness" of only {\MgII} components were tested for significance, whereas for this work, all components from all associated ions were also tested; this resulted in a reduction of the number of components required for the final VP models.  In a survey of {\MgII} absorbers in moderate resolution spectra ($\sim 30$~{\kms}), \citet{petitjean90} found a linear relationship with slope  0.35~{\AA}~cloud$^{-1}$, corresponding to $\simeq\!\!3$~clouds~{\AA}$^{-1}$.  

Overall, we see that the measured VP component line density is strongly affected by the spectral resolution, the signal-to-noise ratio, and the VP fitting approach. It is expected that the higher signal-to-noise ratios and resolutions of the future 30-meter class telescopes will result in an even higher component line density. As such, if any future works undertake a characterization of the component line density, we recommend adopting the approach of enforcing the minimum number of statistically significant components to a well-defined confidence level.

\subsection{VP Component Column Densities}
\label{subsec:ndist}

The VP component column densities are a key input to photoionization models, which are commonly employed to constrain ``cloud" ionization conditions and metallicities, and to explore the spectral energy distribution of the local ionizing radiation field \citep[e.g.,][]{werk14, lehner19, pointon19}. The distribution of column densities also provides key constraints for hydrodynamic cosmological simulations of the circumgalactic and intergalactic medium \citep[e.g.,][]{cwc-direct, oppenhiemer18, peeples19}. The column density distribution obtained from VP decomposition can more effectively account for unresolved saturation in the absorption profiles than direct profile inversion via the apparent optical depth method \citep{savage91, jenkins96}. Furthermore, VP decomposition provides an explicit and well-defined segregation of absorbing components. 

\begin{figure*}[hbt]
\figurenum{8} \centering
\includegraphics[width=0.8\textwidth]{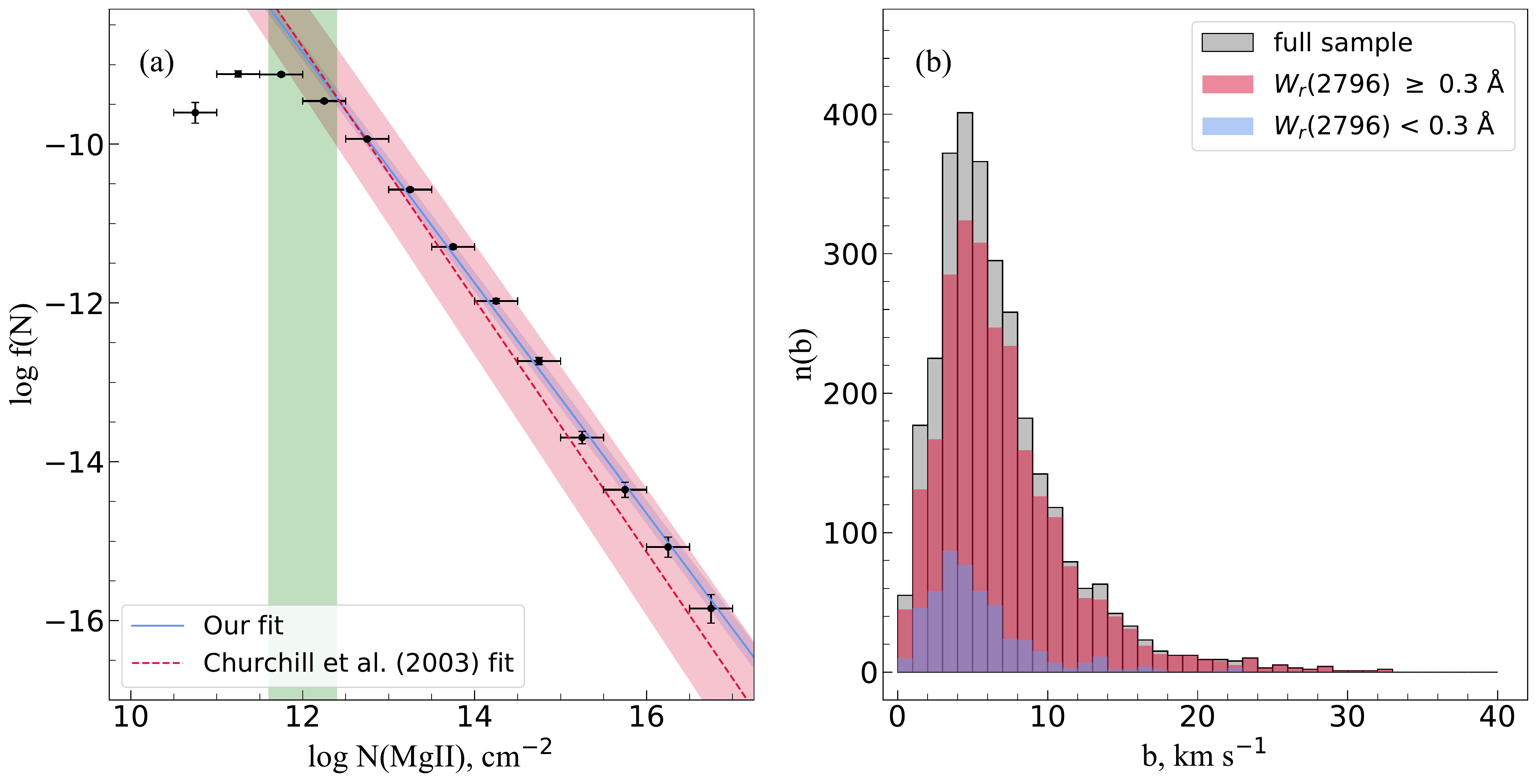}
\caption{(a) The {\MgII} VP column density distribution, $f(N)$.  The vertical pink shaded area indicates the region of partial completeness of the survey (see text).  The fitted line (solid) is the power-law maximum-likelihood fit (on unbinned data), which was performed on all VP column densities of $\log N({\MgII}) > 12.4$ cm$^{-2}$.  The resulting slope is $\delta = 1.45\pm0.01$. {For comparison, the fit (dashed line) and uncertainty (green shading) from \citet{churchill03} is shown.} (b) The binned Doppler $b$ parameter distribution of VP components. The color scheme for the weak system, strong system, and full sample is the same as that in Figure~\ref{fig:zabs}.}
\label{fig:bNdists}
\end{figure*}

VP analysis of our full sample of {\MgII} systems yields the component column density distribution shown in Figure~\ref{fig:bNdists}(a).  The 422 {\MgII} systems comprise a total of 2989 VP components. The distribution has been normalized by this total number of components, so that the quantity $f(N)$ represents the fraction of VP components per unit column density in the sample, a quantity that is reproducible in any survey independent of the number of quasar spectra, absorption line systems, and/or the redshift path sensitivity function of the survey. The pink vertical shaded area indicates the region of partial completeness due to line blending in kinematically complex absorption profiles as determined by the simulations of \citet{churchill03}.  They found that the 90\% completeness levels for unblended and blended lines were log $N$({\MgII}) = 11.6 cm$^{-2}$ and 12.4 cm$^{-2}$, respectively, for spectra having a mean $5\sigma$ equivalent width sensitivity of $W_r = 0.02$ {\AA}. ``Completeness level'' refers to the percentage of simulated components of a given column density recovered during VP analysis. In the shaded region on Figure~\ref{fig:bNdists}(a), components in complex profiles can be lost due to blending, though the completeness for unblended components is 90\%.  Below this region, even unblended (single component) absorbers can be lost due to the signal-to-noise ratio of the spectra.

The column density distribution can be fit by a power law, 
\begin{equation}
f(N) = CN^{-\delta},
\end{equation}
where $f(N)$ is the fraction of clouds with column density $N$ per unit column density, $C$ is a normalization constant, and $\delta$ is the power law slope. The maximum likelihood method was used to obtain the power law fit to the unbinned data \citep[see][]{churchill97}.  The fit was performed only on column densities above the region of partial completeness ($\log N({\MgII}) > 12.4$~cm$^{-2}$) so as to not skew the slope.  We obtained  $\delta = 1.45\pm0.01$.  In a study of 14 {\MgII} systems containing 33 VP components, \citet{petitjean90} obtained a significantly shallower slope of $\delta = 1.0\pm0.1$, although their spectral resolution was lower ($\simeq 30$~{\kms}).  {In a study with identical resolution, \citet{churchill03} obtained $\delta = 1.59\pm0.05$ by fitting their sample of 175 VP components in 23 {\MgII} systems.  Our slightly shallower slope is likely due to the modifications we made to {\sc Minfit}, which resulted in a lower VP component line density (see Section~\ref{subsec:nvp}) skewed slightly toward higher column density components.  }

\subsection{VP Component Doppler Parameters}
\label{subsec:bdist}

In Figure~\ref{fig:bNdists}(b), we plot the {\MgII} Doppler $b$ parameter distribution of the VP components. The median Doppler $b$ parameters and standard deviations are $4.5\pm3.5$~{\kms}, $6.0\pm4.5$~{\kms}, and $5.7\pm4.4$~{\kms} for the weak (blue), strong (pink), and full (gray) samples, respectively.  \citet{churchill97} found a median Doppler parameter of $\sim\!\! 3.5$ {\kms} in a study of 48 {\MgII} systems and \citet{churchill03} found $5.4\pm4.3$ in a study of 23 {\MgII} systems; in both cases the data were of comparable quality and resolution (6.6~{\kms}). \citet{petitjean90} found in their study of 14 {\MgII} systems that the $b$ distribution peaked between 10--15 {\kms}; however, their larger value was because their data were of lower spectral resolution (30~{\kms}) and they noted that this significantly distorted the observed distribution. 

Based on simulations designed to test the recovery of the Doppler $b$ distribution in HIRES spectra \citep{churchill03}, the observed distribution peak is likely $\sim\!\! 1$--2 {\kms} too high relative to the true underlying distribution (assuming {\MgII} absorption profiles arise in spatially separated isothermal clouds giving rise to Voigt profiles).  In addition, the distribution tail at high $b$ values has been shown in these simulations to be an artifact of component blending and unresolved saturation.  As mentioned in Section~\ref{subsec:nvp}, $\sim\!\!30$\% of {\it simulated components} are not recovered in the VP decomposition for our spectral quality.  As a result, some $b$ parameters in the observed distribution are too broad compared to the ``true" distribution.

If the Doppler broadening is assumed to be predominately thermal, then the observed $b$ parameter distribution medians correspond to gas temperatures of $\sim\!\! 30,000$~K, $\sim \!\! 53,000$~K, and $\sim\!\! 47,000$~K for the weak, strong, and full samples, respectively. However, applying the 1--2~{\kms} correction to the mode of the $b$ parameter distribution would produce a median temperature, in the case of the weak sample, of $\sim\!\! 9000$--18,000~K, in strong sample, of $\sim\!\! 23,000$--37,000~K, and for full sample, a median temperature of $\sim\!\! 20,000$--32,000~K. 


Based on simulations designed to test the recovery of the Doppler $b$ distribution in HIRES spectra \citep{churchill03}, the observed distribution peak is likely $\sim\!\! 1$--2 {\kms} too high relative to the true underlying distribution (assuming {\MgII} absorption profiles arise in spatially separated isothermal clouds giving rise to Voigt profiles).  In addition, the distribution tail at high $b$ values has been shown in these simulations to be an artifact of component blending and unresolved saturation.  As mentioned in Section~\ref{subsec:nvp}, $\sim\!\!30$\% of {\it simulated components} are not recovered in the VP decomposition for our spectral quality.  As a result, some $b$ parameters in the observed distribution are too broad compared to the ``true" distribution.

If the Doppler broadening is assumed to be predominately thermal, then the observed $b$ parameter distribution medians correspond to gas temperatures of $\sim\!\! 30,000$~K, $\sim \!\! 53,000$~K, and $\sim\!\! 47,000$~K for the weak, strong, and full samples, respectively. However, applying the 1--2~{\kms} correction to the mode of the $b$ parameter distribution would produce a median temperature, in the case of the weak sample, of $\sim\!\! 9000$--18,000~K, in strong sample, of $\sim\!\! 23,000$--37,000~K, and for full sample, a median temperature of $\sim\!\! 20,000$--32,000~K.

\subsection{VP Component Velocity Clustering}
\label{sec:tpcf}

\begin{figure}[thb]
\figurenum{9} \epsscale{1.1} 
\plotone{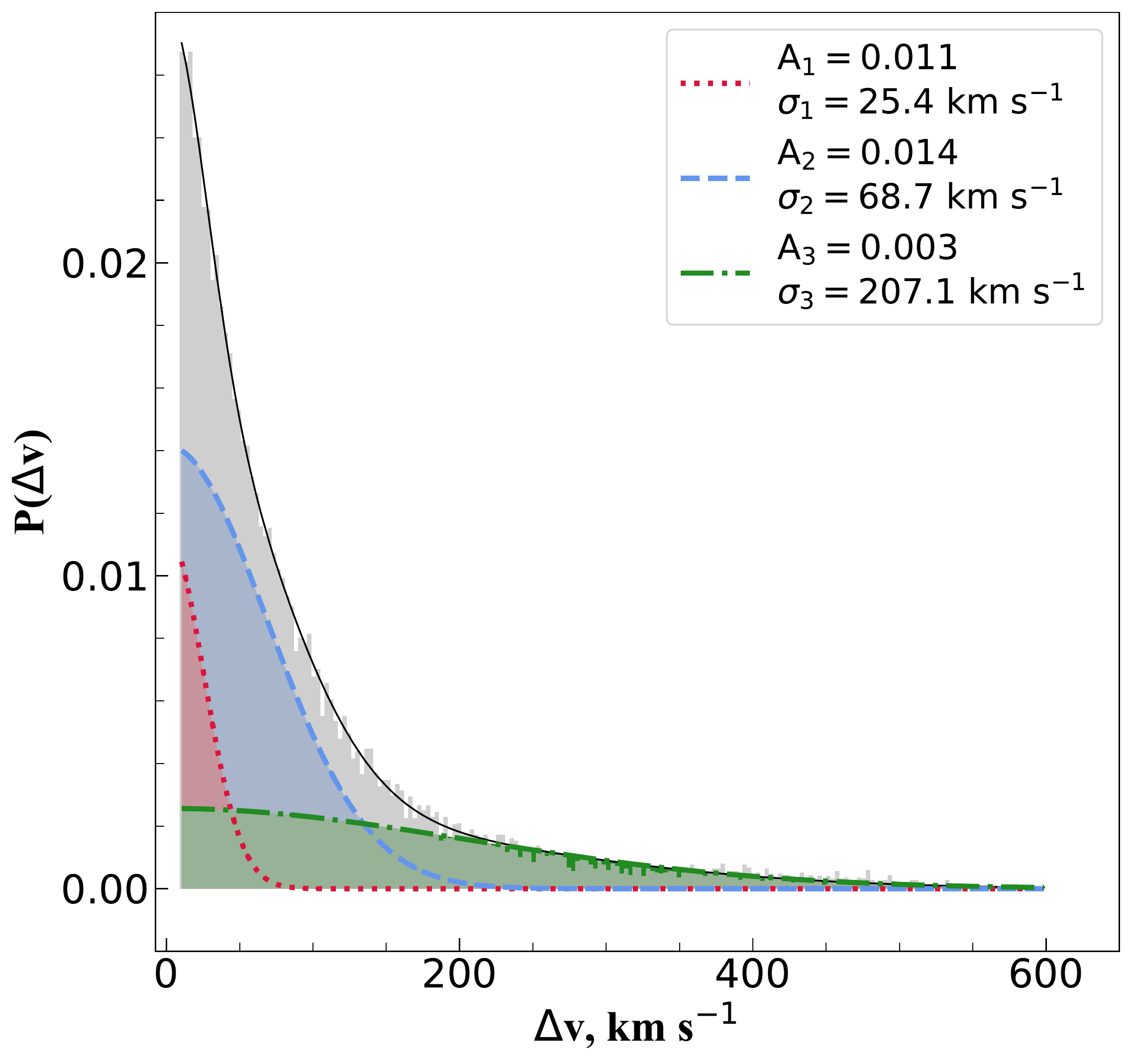} 
\caption{The VP component velocity two-point correlation function (TPCF) for the full sample is shown in gray; velocity separations $\Delta v$ are plotted in 3 {\kms} bins.  The probability of finding two VP components separated by $\Delta v$ has been modelled with a three-component composite Gaussian.  The dotted, dashed and dot-dashed curves represent components 1 (pink), 2 (olive), and 3 (green) of the composite Gaussian, respectively; the black solid curve is the total. Component pair splittings with $\Delta v < 10$ {\kms} are omitted. The $A$ values represent the Gaussian component amplitudes, and the $\sigma$ values represent the Gaussian component velocity dispersions. The binned data (gray) are presented in Table~\ref{tab:tpcf}.}
\vglue 0.05in
\label{fig:tpcf}
\end{figure}

\begin{deluxetable}{rcrcr}
\tablewidth{\linewidth}
\tabletypesize{\normalsize}
\tablecaption{TPCF Data \label{tab:tpcf}}
\tablehead{
\colhead{$\Delta v$\tablenotemark{a}} & \colhead{} &
\colhead{$P(\Delta v)$} & \colhead{} &
\colhead{$\sigma(P)$} \\[-4pt]
\colhead{[\kms]} & \colhead{} &
\colhead{$10^{-2}$} & \colhead{} &
\colhead{$10^{-2}$}
}
\startdata
1.5\tablenotemark{b} & & 1.619 & & 0.077\\[-2pt]
4.5\tablenotemark{b} & & 1.519 & & 0.075\\[-2pt]
7.5\tablenotemark{b} & & 2.014 & & 0.087\\[-2pt]
10.5 & & 2.468 & & 0.096\\[-2pt]
13.5 & & 2.438 & & 0.095\\[-2pt]
16.5 & & 2.466 & & 0.096\\[-2pt]
19.5 & & 2.215 & & 0.091\\[-2pt]
22.5 & & 2.212 & & 0.090\\[-2pt]
25.5 & & 2.010 & & 0.086\\[-2pt]
28.5 & & 2.003 & & 0.086\\
\enddata
\tablenotetext{a}{The bin sizes are 3~{\kms}. These values are the bin centers.}
\tablenotetext{b}{$P(\Delta v)$ affected by component blending.}
\tablecomments{Table~\ref{tab:tpcf} is published in its entirety in machine-readable format. A portion is shown here for guidance regarding its form and content.}
\end{deluxetable}

The velocity clustering of the VP components is quantified using the velocity two-point correlation function \citep[TPCF,][]{petitjean90}. The TPCF is the probability, $P(\Delta v)$, that any randomly selected pair of VP components within a system will have a velocity separation $\Delta v$. {The velocity TPCF for our full sample is shown in Figure~\ref{fig:tpcf} and tablulated in Table~\ref{tab:tpcf}.} This probability distribution is commonly fit with a composite Gaussian function,
\begin{equation}
 P(\Delta v) = \sum_{n=1}^{N} g_n(\Delta v) \, ,
\end{equation}
where $N$ is the number of components, 
\begin{equation}
g_n(\Delta v) = \frac{1}{\sqrt{2\pi}}\frac{a_n}{\sigma_n}
\exp \left\{-\frac{(\Delta v)^2}{2\sigma_n^2} \right\}\, , 
\end{equation}
are the Gaussian function components, and where $a_n$ is a fitted scaling factor and $\sigma_n$ is a fitted velocity dispersion.  The amplitude of each component is then
\begin{equation}
A_n = \frac{1}{\sqrt{2\pi}}\frac{a_n}{\sigma_n} \, .
\end{equation}
\citet{petitjean90} and \citet{churchill03}, in their studies of 14 and 23 {\MgII} absorption systems, respectively, fit their TPCF distributions of VP components using two-component Gaussian models. The results of \citet{petitjean90} were $\sigma_1 = 80$~{\kms} and $\sigma_2 = 390$~{\kms} for a spectral resolution of 30~{\kms}.  The authors attributed the narrower of these two components to motions within galaxy halos, and the broader to the motions of galaxy pairs.  However, the \citet{churchill03} study, which had 6.6~{\kms} velocity resolution, calculated best fit dispersion of $\sigma_1 = 54$~{\kms} and $\sigma_2 = 166$~{\kms}. They suggested that the {\MgII} component velocity dispersion might reflect the range observed in face-on galaxy disks and edge-on galaxy rotational motions, as well as infall and outflow in the halos. The larger TPCF dispersion reported by Petitjean \& Bergeron \nocite{petitjean90} were likely due to resolution effects that prevented identification of smaller VP component velocity splittings.

We fitted the TPCF from our full sample with a three-component composite Gaussian function; the resulting functions are superimposed on the TPCF in Figure~\ref{fig:tpcf}.  Three components were used because two did not adequately fit the extended tail of our distribution.  Our resulting velocity dispersions are $\sigma_1 = 25.4$~{\kms}, $\sigma_2 = 68.7$~{\kms}, and $\sigma_3 = 207.1$~{\kms}.  Velocity separations of $\Delta v < 10$~{\kms} were excluded from the fit because their relative numbers are artificially lowered due to component blending at small $\Delta v$.

Small velocity separations are the most probable. The probability drops steeply up to $\Delta v \sim 150$~{\kms}; at larger separations the probability decreases more slowly to our maximum veocity separation of $\Delta v = 572.8$~{\kms}. Considering the modern view of the kinematics of the low-ionization circumgalactic medium \citep[e.g.,][]{weiner09,kacprzak10, martin12, nielsen15, nielsen16, ho17, kacprzak17, zabl19, zabl20}, it would be a gross over interpretation of the TPCF parameterization to identify each Gaussian component with a specific galactic kinematic component or physical phenomenon related to the circumgalactic baryon cycle. Parameterizing the TPCF by a functional fit provides a convenient functional characterization of VP component velocity clustering. One should be cautious to consider the equivalent width detection threshold and the spectral resolution when comparing the TPCF.  We remind the reader that we applied a uniform detection threshold criteria for a absorbing system to be included in our analysis of kinematics (see Section~\ref{subsec:kinematic}). Thus criteria ensures that the kinematics analysis has a uniform sensitivity to small equivalent width absorption features at high velocities from system to system.

\section{Conclusion}
\label{sec:conclude}

We searched 249 HIRES and UVES quasar spectra and identified 480 {\MgII} absorbers in 186 of the quasar lines of sight. The full sample spans the  equivalent width range  $0.006 \leq W_r(2796) \leq 6.23$ {\AA} over the redshift range $0.19 \leq z \leq 2.55$, with a mean of $\langle z \rangle = 1.18$. We present the absorption properties of the complete sample in Table~\ref{tab:samplesysanaldata}.  

We compared the equivalent width distribution of the complete sample with that of the unbiased survey of \citet{nestor05}, and found that our sample is not inconsistent with being a fair sample, though we have a slight overabundance of $W_r(2796)\! \sim\! 2$~{\AA} systems. We thus proceed under the assumption that our sample is a fair sample for studying the kinematics of the {\MgII} systems.  

In this paper, we examined and present the global kinematic properties of the {\MgII} absorbers. The kinematics of the systems were quantified using the formalism of Voigt profile (VP) fitting. We employed the program {\sc Minfit} \citep{churchill97}. The majority of the fitting comprises the doctoral thesis research of \citet{evans-phd}. For the kinematic analysis, we limited our study to the ``kinematic sample'', i.e., those absorbers which have a $5\sigma$ detection threshold of $W_r(2796) \leq 0.02$~{\AA} across a velocity window of $\pm 600$~{\kms} centered on the {\MgII}~$\lambda 2796$ profiles (see Section~\ref{sec:absorptionchars}).  The kinematic sample comprises 422 systems found in 163 of the quasar spectra.  Based on historical precedent, we classified 180 of these absorbers as weak systems (having $W_r(2796) < 0.3$ {\AA}) and 242 as strong systems (having $W_r(2796) \geq 0.3$ {\AA}). The VP fitting yielded a total of 2989 components, with an average of 2.7 and 10.3 components being recovered for the weak and strong {\MgII} subsamples, respectively.  The VP fitting parameters of the kinematic sample are presented in Table~\ref{tab:sampleminfitnumbers}. \\

Key quantitative results are:

\begin{enumerate}

\item We find a VP component line density of $8.62 \pm 0.23$  clouds {\AA}$^{-1}$.  Fitting our VP component column density distribution over the range $12.4 \leq$ log $N({\MgII}) \leq 17.0$ cm$^{-2}$ resulted in a power law slope of $\delta = 1.45\pm0.01$.

\item Examining the {\MgII} Doppler $b$ parameter distribution of the VP components, we find that the median Doppler $b$ parameters are $4.5\pm3.5$~{\kms}, $6.0\pm4.5$~{\kms}, and $5.7\pm4.4$~{\kms} for the weak, strong, and full samples, respectively. These medians, after correcting for the 1--2~{\kms} correction from the simulations, imply gas temperatures of $T \sim\!\! 9000$--18,000~K for the weak systems, $T \sim\!\! 23,000$--37,000~K for the strong systems, and $T \sim\!\! 20,000$--32,000~K for full sample. 

\item We modeled the probability of component velocity splitting (the two-point velocity correlation function, TPCF) of our full sample using a three-component composite Gaussian function. Our resulting velocity dispersions are $\sigma_1 = 25.4$~{\kms}, $\sigma_2 = 68.7$~{\kms}, and $\sigma_3 = 207.1$~{\kms}.  Though we so not assign a physical or kinematic component of galaxies or the the CGM to each Gaussian component, we would surmise that the low amplitude, high velocity tail of the TPCF might be associated with outflows in galaxies with active star formation \citep{nielsen15}.

\end{enumerate}

Future work with the data presented would include studying cosmic evolution in the {\MgII} absorber kinematics, photoionization modeling of the absorbers to constrain absorber metallicities, cloud sizes, and masses, and ionization conditions, including effects of the ultraviolet background evolution. As the quasar spectra comprising this sample do not include the KODIAQ data releases \citep{omeara15, omeara17}, nor the complete data release of the UVES SQUAD \citep{uves-squad}, there remains the opportunity to increase the sample size.  This would be essential for studying the redshift evolution of {\MgII} absorber kinematics, both directly from the flux decrements, and using the VP fitting parameters.

\begin{center}
ACKNOWLEDGMENTS
\end{center}

We dedicate this paper to memory of Dr.\ Wallace Leslie William Sargent, who was a pioneer of the field of quasar absorption lines and so positively influenced the lives and careers of multiple generations of astronomers.  We thank Wallace Sargent, Michael Rauch, J.~Xavier Prochaska, and Charles Steidel for their contribution of HIRES spectra from the pre-WKMO archival period.  We thank the members of the UVES SQUAD who contributed to the first data release. This research has made use of the services of the ESO Science Archive Facility. 
Some of the data presented herein were obtained at the W. M. Keck Observatory, which is operated as a scientific partnership among the California Institute of Technology, the University of California, and the National Aeronautics and Space Administration and made by possible by support of the W. M. Keck Foundation. The authors wish to recognize and acknowledge the very significant cultural role and reverence that the summit of Maunakea has always had within the indigenous Hawaiian community.  We are most fortunate to have the opportunity to conduct observations from this mountain. CWC is grateful for NSF grant AST 0708210, the primary funding for this work; JLE was also supported by a three-year Aerospace Cluster Fellowship administered by the Vice Provost of Research at New Mexico State University and by two-year New Mexico Space Grant Graduate Research Fellowship. Parts of this research were supported by the Australian Research Council Centre of Excellence for All Sky Astrophysics in 3 Dimensions (ASTRO 3D), through project number CE170100012. MTM thanks the Australian Research Council for a QEII Research Fellowship (DP0877998).

\appendix

\section{Notes on Individual Systems}
\label{sec:appendix}

Notes on individual systems are published as electronic material in the online version of the journal article.

\end{document}